\documentclass[onecolumn,aps,pre,preprint,groupedaddress,showpacs,floatfix]{revtex4-1}
\usepackage{graphicx, caption, subcaption}
\usepackage{epsfig}
\usepackage{amsfonts}
\usepackage{color}
\usepackage{ulem}
\usepackage{amsmath}
\usepackage{mathrsfs}
\usepackage[autostyle]{csquotes}
\usepackage{bigints}
\usepackage{floatrow}
\usepackage{caption}
\usepackage{soul}
\usepackage{lipsum}
\usepackage{marginnote}
\definecolor{brown}{rgb}{0.65,0.16,0.16}

\date{today}

\begin{document}
	
\title{Stochastic Stabilization of Transient Axisymmetric Taylor-Couette Flow}
\author{Larry E. Godwin}
\affiliation{Mathematics, Systems Analytics Research Institute, Aston University, Birmingham B4 7ET, England}
\author{Sotos C. Generalis}
\affiliation{Mathematics, Systems Analytics Research Institute, Aston University, Birmingham B4 7ET, England}
\author{Amit K. Chattopadhyay}
\affiliation{Mathematics, Systems Analytics Research Institute, Aston University, Birmingham B4 7ET, England}
\email{a.k.chattopadhyay@aston.ac.uk}

\begin{abstract}
Structured on the paradigmatic Navier-Stokes flow model, we study a stochastically forced Taylor-Couette system in the narrow gap limit, in order to analyze the simultaneous impact of a non-conserved (Gaussian) force and a nonlinear perturbation, in determining linear stability. Our analysis identifies key parametric windows within which the model remarkably retains its linear stability, even against jointly varying stochastic forcing and nonlinear fluctuations. We identify this feature as a {\it latent universality}, that we then utilize to answer the elusive question as to how a recent groundbreaking accretion flow model retains its stability, even in presence of nonlinear perturbations and non-conserved stochastic forcing. Our analytical outline goes beyond the immediate model studied and lays a generic foundation of analyzing stability for nonlinear sheared models acted on by external forces and subjected to boundary layer instability. \\

\end{abstract}
\date{\today}


\maketitle

\newpage

\section{Introduction}
Stochastically  forced Taylor-Couette  flow  represents  the  simplest  macroscopic  nonequilibrium  system,  in  which  the enforced processes have origins at the molecular level, with instabilities driven by molecular fluctuations. The mean flow Reynolds number ($Re$) provides a single, control and combined parameter to measure the departure of the fluid from its equilibrium state. Investigation of this simple macroscopic nonequilibrium system is of interest from both physical and \enquote{local}, cf. hydrodynamic turbulence,  standpoints. The hydrodynamic equations describe the macroscopic flow with high accuracy \cite{agb}. 

The microscopic description of fluid  mechanics  has  not  been  fully established.  The investigation of this simplest  flow  system contributes  to eliminating this  deficiency.  Additionally, during the last decade, the microscopic description of fluid  mechanics has been under intense focus. The formalism has been found to be particularly useful in understanding  the perturbed dynamics of the transient mean flow energy that was always known to exist despite the exponential stability of all normal modes of the system. As initially indicated in \cite{Trefethen1993,Reddy93,Tahia06} and later established in \cite{bani,tc02}, the  transient growth of  perturbations was attributed to the non-normal growth of the linearized dynamical operators of the shear flow systems. A stochastic interpretation of this effect, specifically for boundary layer turbulence, modeled as the sheared Taylor-Couette flow of a viscous fluid confined in between two coaxial rotating cylinders, was analysed through a series of publications \cite{akc1,akc2,akc3} based on a theoretical foundation laid down in \cite{akc4}. 

\begin{center}
\begin{figure}[tbp]
	\includegraphics[width=50mm]{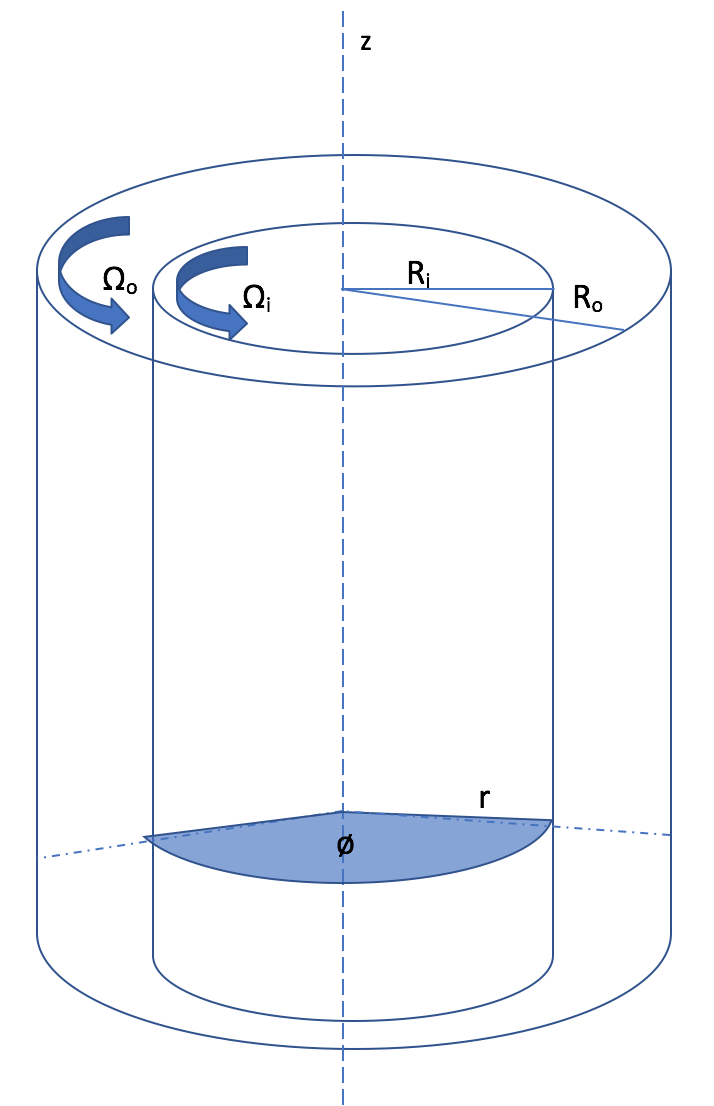}
	\caption{Schematic diagram of the Taylor-Couette Flow System}
	\label{fig01}
\end{figure}
\end{center}

The sheared flow profile analyzed in these works \cite{bani,tc02,akc1,akc2,akc3} is depicted in Fig. \ref{fig01}. The flow lines curling around the axial symmetry line (dashed line) clearly indicate increasing perturbations closer to the boundary layer. It is well known from previous works (see, for example \cite{tc10}) that in the small gap limit, the onset of instability of the basic axisymmetric solution for nearly co-rotating cylinders occurs for $R=\frac{ 106.8}{ \Omega}+\Omega$. There the two co-axial cylinders with radii $r_1$ and $r_2$, respectively rotate with speeds $\Omega_1 r_1$ and $\Omega_2 r_2$ leading to 
$\Omega=\left(\Omega_1+\Omega_2\right)d^2/\nu$, where $\Omega$ is twice the mean rotation rate in dimensionless units and $d$ is the width.
Expressed as a function of the inner and outer Reynolds numbers ${Re}_i$ and ${Re}_o$, and their respective rotational frequencies, the Rayleigh line and the linear stability boundary jointly define the linear stability (or instability) regimes of such flows \cite{tc02,bani}.The azimuthal independent solution bifurcating from the basic, undisturbed state, is characterized by the critical value $\beta_c=1.558$ of the azimuthal wavenumber $\beta$, and is called Taylor vortex flow (Tvf). As this solution evolves as a function of the parameters $R$ and $\Omega$, it encounters stability boundaries corresponding to the onset of three-dimensional disturbances of the various kinds. 

While transition from a laminar to a turbulent regime is a well studied feature of the Taylor-Couette flow \cite{tc10}, the issue of fast non-normal transient growth of a linearly stable laminar flow profile, leading to strong instability, \cite{tc01,bani} is a much less studied topic. As demonstrated in \cite{tc01,bani,akc1,akc2}, such subcritical sheared transition, that is transition to turbulence notwithstanding linear stability, is driven by uncontrolled columnar growth and is not a theoretical artefact, rather an inherent (laminar) flow instability resulting from nonnormal eigenmodes. The exponentially growing azimuthal and axial velocities define a nonnormal basis \cite{tc02} that leads to selective growth of eigenmodes \cite{Butler1992, Trefethen1993}. This understanding successfully explained a key unsolved problem concerning magnetorotational instability induced transport in cold accretion disks at temperatures below 3000 $^\circ$K.

A landmark work in the physics of linear Magneto-Hydro-Dynamic (MHD) turbulence is the paper by Balbus and Hawley \cite{Balbus1991} who showed that magnetorotational instability can interact with the Chandrasekhar instability \cite{Chandra1960} to create MHD turbulence. However, this line of argument fell short when such instabilities were shown to exist even for non-magnetic accretion flows. 
In three successive key publications, Chattopadhyay and coworkers showed that thermal fluctuation fields could accomplish the equivalent of non-normal transport \cite{akc1,akc2,akc3} that then explained the so-called {\it cold accretion turbulence} while retaining the ubiquitous Kolmogorov scaling intact. The present work offers the first unified theoretical description combining non-normal deterministic turbulence with stochastically driven instabilities. 

While transition from a laminar to a turbulent regime is a well studied feature of the Taylor-Couette flow \cite{tc10}, the issue of fast non-normal transient growth of a linearly stable laminar flow profile, leading to strong instability, \cite{tc01,bani} is a much less studied topic. As demonstrated in \cite{tc01,bani,akc1,akc2}, such subcritical sheared transition, that is transition to turbulence notwithstanding linear stability, is driven by uncontrolled columnar growth and is not a theoretical artefact, rather an inherent (laminar) flow instability resulting from nonnormal eigenmodes. The exponentially growing azimuthal and axial velocities define a nonnormal basis \cite{tc02} that leads to selective growth of eigenmodes \cite{Butler1992, Trefethen1993}. This understanding successfully explained a key unsolved problem concerning magnetorotational instability induced transport in cold accretion disks at temperatures below 3000 $^\circ$K. 
A landmark work in the physics of linear Magneto-Hydro-Dynamic (MHD) turbulence is the paper by Balbus and Hawley \cite{Balbus1991} who showed that magnetorotational instability can interact with the Chandrasekhar instability \cite{Chandra1960} to create MHD turbulence. However, this line of argument fails to explain why such instabilities exist even for non-magnetic accretion flows. In three landmark publications, Chattopadhyay and coworkers showed that thermal fluctuation fields could accomplish the equivalent of non-normal transport \cite{akc1,akc2,akc3} that then explained the so-called {\it cold accretion turbulence} while retaining the ubiquitous Kolmogorov scaling intact. 

The present work offers the first unified theoretical description combining non-normal deterministic turbulence with stochastically driven instabilities. This paper is organized as follows. Section II provides the mathematical foundation of the problem and derives the stochastically driven linearized Taylor-Couette flow amplitude equations. Section III exposes the numerical scheme employed to derive the eigenvalue spectrum of the amplitude equations of section II. In section IV, we discuss our results, where we also present the numerical simulation that supports bifurcation characteristics obtained on a weakly nonlinear basis. Conclusions and prospective outliers are outlined in section V.

\section{Flow Model}
In line with the original MHD perturbed model \cite{akc1,akc2} for an incompressible Navier-Stokes flow, we assume a flow velocity $\bf v$ for a Newtonian fluid with kinematic viscosity $\nu$ that is sandwiched between two infinite rotating cylinders with inner and outer radius $r^*_i$ and $r^*_o$ respectively. The gap width $d$ scales as the radius difference and is given by $d = r^*_o - r^*_i$; the viscous time scales as $\nu^{-1} d^2$, with a non-dimensional pressure proportional to $\nu^{-2} d^2$. The (nondimensional) reduced pressure is given by $\tilde{p} = \rho^{-1}\mathrm{p}$, leading to the flow model as follows:

\begin{equation}
\label{gov1}
\frac{\partial \bf v}{\partial t} = (\bf{v}.\nabla)\bf{v} - \bf{\nabla}\tilde{p} + \nabla^2 \bf{v}, 
\end{equation}

\begin{equation}
\label{gov2}
\nabla \cdot \bf{v} = 0.
\end{equation}

\noindent
In cylindrical coordinates, the velocity $\bf{v}$ can be defined as:
\begin{equation}
\label{gov3}
{\bf{v}} = v_r{\bf{e}}_r + v_\theta{\bf{e}}_\theta + v_z {\bf{e}}_z ,
\end{equation}
where the orthonormal components are defined as follows: radial: $({\bf e_r})$ = $(1, 0, 0)^T$, azimuthal: $({\bf e_\theta})$ = $(0, 1, 0)^T$ and axial: $({\bf e_z})$ = $(0, 0, 1)^T$. It is assumed that the basic flow is independent of time $t$, as also of the azimuthal and axial directions; it is also considered translationally invariant along the radial direction, the usual theoretical architecture employed to deal with reflection symmetry due to the walls of the cylinders. The no-slip boundary condition is then defined as:
\begin{equation}
\label{gov4}
{\bf{v}}|_{r = r_i} = Re_i\:{\bf{e}_i} \quad \text{and} \quad  {\bf{v}}|_{r = r_o} = Re_o\:\bf{e}_o.
\end{equation}

\noindent
Equation (\ref{gov4}) is combined with Eq. (\ref{gov1}) and the incompressibility condition in Eq. (\ref{gov2}) to define the constrained circular Taylor-Couette flow as follows:
\begin{equation}
\label{gov5}
{\bf{v}}^B = \left(Ar + \frac{B}{r} \right)\ {\bf{e}}_\theta  \quad \text{and} \quad p^B = \frac{1}{2}A^2r^2 + 2AB\ln(r) - \frac{B^2}{2r^2}.
\end{equation}

\noindent
Here the constants A and B are given by:
\begin{equation}
\label{gov6}
A = \frac{Re_o - \eta Re_i}{1 + \eta} \quad \text{and} \quad B = \frac{\eta(Re_i - \eta Re_o)}{(1 - \eta)(1 - \eta^2)}.
\end{equation}

\noindent
The description above uses non-dimensional parameters $r_i=r_i^*/d$ (scaled inner radius), $r_o=r_o^*/d$ (scaled outer radius), $\eta=r_i/r_o$ (radius ratio), ${Re}_i=\frac{d}{\nu}r_i^* \Omega_i$ (inner Reynolds number) and ${Re}_o=\frac{d}{\nu}r_o^* \Omega_o$.
The Navier-Stokes (NS) equation is then linearized around the symmetric mean flow. The flow solutions from the NS-model converge to zero everywhere within the interval $r \in [r_i, r_o]$ apart from the azimuthal direction:
\begin{equation}
\label{gov7}
v_r = 0, \quad v_\theta = Ar + \frac{B}{r}, \quad \text{and} \quad v_z = 0.
\end{equation}

\noindent
The combined linerarized description from eqns. (\ref{gov1}) and (\ref{gov2}) then shows up as 
\begin{equation}
\label{gov8}
\frac{\partial {\bf v}}{\partial t} = -({\bf{v}}^B \cdot {\bf{\nabla}}){\bf{u}} - ({\bf{u}} \cdot {\bf{\nabla}}){\bf{v^B}}  - {\bf{\nabla}}{\tilde{p}} + \nabla^2 \bf{v}.
\end{equation}

\subsection{Linearly Stable Model}
In order to analyze the stability of the Taylor-Couette flow, the circular Couette flow solution is perturbed around its fixed point with Fourier (harmonic) components \cite{Butler1992,tc02}, where we assume periodicity in both axial and azimuthal directions:
\begin{eqnarray}
\label{gov9}
{\bf{v}} & = & {\bf{v}^B  + \bf{u}(r)}e^{i(\alpha z + \beta \theta + \lambda t)} +\text{complex\:\:conjugate},\\
{\bf{\tilde{p}}} & = & {\bf{p^B  + p(r)}}e^{i(\alpha z + \beta \theta + \lambda t)}  + \text{complex\:\:conjugate}.
\end{eqnarray}

\noindent
Here the axial and azimuthal wave numbers are  $ \alpha \in \mathbb{R} $, and $ \beta \in \mathbb{Z} $, respectively and $ \lambda \in \mathbb{C} $. The velocity $\bf{u(r)}$ needs to satisfy both the boundary and divergence conditions:
\begin{equation}
\label{gov10}
\bf{u(r_i)} = \bf{u(r_o)} = 0 \quad and \quad  \nabla \cdot \bf{u} = 0
\end{equation}

\noindent
Using cylindrical coordinates, the linearized Navier-Stokes Eq. (\ref{gov8}) is then solved using the Fourier ansatz as in Eq. (\ref{gov9}), discarding all higher-ordered  nonlinear terms. This leads to a spatially three-dimensional eigenvalue problem for the axial ($\alpha$) and azimuthal ($\beta$) modes defining the perturbed system. The momentum and the continuity equations then become:
\begin{equation}
\label{gov11}
\lambda u_r  =  \Bigg[D_+D - \frac{\beta^2 + 1}{r^2} - \alpha^2 - \frac{i\beta}{r}v_\theta^B\Bigg] u_r + \Bigg[\frac{2}{r}v_\theta^B - \frac{2in}{r^2}\Bigg]u_\theta + Dp,
\end{equation}

\begin{equation}
\label{gov12}
\lambda u_\theta  =  \Bigg[D_+D - \frac{\beta^2 + 1}{r^2} - \alpha^2 - \frac{i\beta}{r}v_\theta^B\Bigg] u_\theta + \Bigg[\frac{2in}{r^2} - 2A\Bigg]u_r + \frac{i\beta}{r}p,
\end{equation}

\begin{equation}
\label{gov13}
\lambda u_z  =  \Bigg[D_+D - \frac{\beta^2 }{r^2} - \alpha^2 - \frac{i\beta}{r}v_\theta^B\Bigg] u_z  +  i\alpha p,
\end{equation}

\begin{equation}
\label{gov14}
D_+u_r + \frac{i\beta}{r} u_\theta + i\alpha u_z = 0.
\end{equation}

\noindent
The linear differential operators $D = \frac{d}{dr}$ and $D_+ =  D + \frac{1}{r}$ used above were first introduced by Chandrasekhar \cite{tc10}. Equations (\ref{gov11}-\ref{gov14}) can be unified in a matrix representation as follows:
\begin{equation}
\label{gov15}
\lambda  \bf{u} =  \bf{\mathcal{L}} \bf{u}  + \bf{\zeta} p,
\end{equation}

\noindent
where the (cylindrical) components of the linear operator $\bf{\mathcal{L}}$ are described below:
\begin{equation}
\label{gov16}
\begin{aligned}
\mathcal{L}_{rr} = \mathcal{L}_{\theta \theta} &= D_+D - \frac{\beta^2 + 1}{r^2} - \alpha^2 - \frac{i\beta}{r}v_\theta^B \\
\mathcal{L}_{r \theta}  &= \frac{2}{r}v_\theta^B - \frac{2i\beta}{r^2} \\
\mathcal{L}_{\theta r}  &= \frac{2i\beta}{r^2} - 2A \\
\mathcal{L}_{zz} &= \mathcal{L}_{rr} + \frac{1}{r^2} .
\end{aligned}
\end{equation}
The resultant equation in its matrix form is then shown as follows:
\begin{equation}
\label{gov17}
\lambda
\begin{bmatrix}
u_r \\
u_\theta\\
u_z \\
\end{bmatrix}
=
\begin{bmatrix}
\mathcal{L}_{rr}          & \mathcal{L}_{r \theta}          & 0  \\
\mathcal{L}_{\theta r} & \mathcal{L}_{\theta \theta}  & 0  \\
0                     &                0                          & \mathcal{L}_{zz}  
\end{bmatrix}
\begin{bmatrix}
u_r \\
u_\theta\\
u_z 
\end{bmatrix} 
+ 
\begin{bmatrix}
D \\
\frac{i\beta}{r}\\
i\alpha\\
\end{bmatrix}p.
\end{equation}

\subsection {Stochastically driven linearized Taylor-Couette Flow}
The role of stochasticity in dynamical systems is an often quoted topic with wide applications \cite{tc15}. Analyzing fluid flows in constrained geometries, uncertainty or randomness may arise due to energy amplification of potentials caused by stochastic excitations of the operator $\bf{\mathcal{L}}$ defining the system dynamics. For instance, fluctuations in the basic flow, typically perturb the steady flow profile of the Taylor-Couette geometry, leading to fluctuations in the free flow of the particles of the fluid.
\begin{center}
\begin{figure}[tbp]
	\includegraphics[width=170mm]{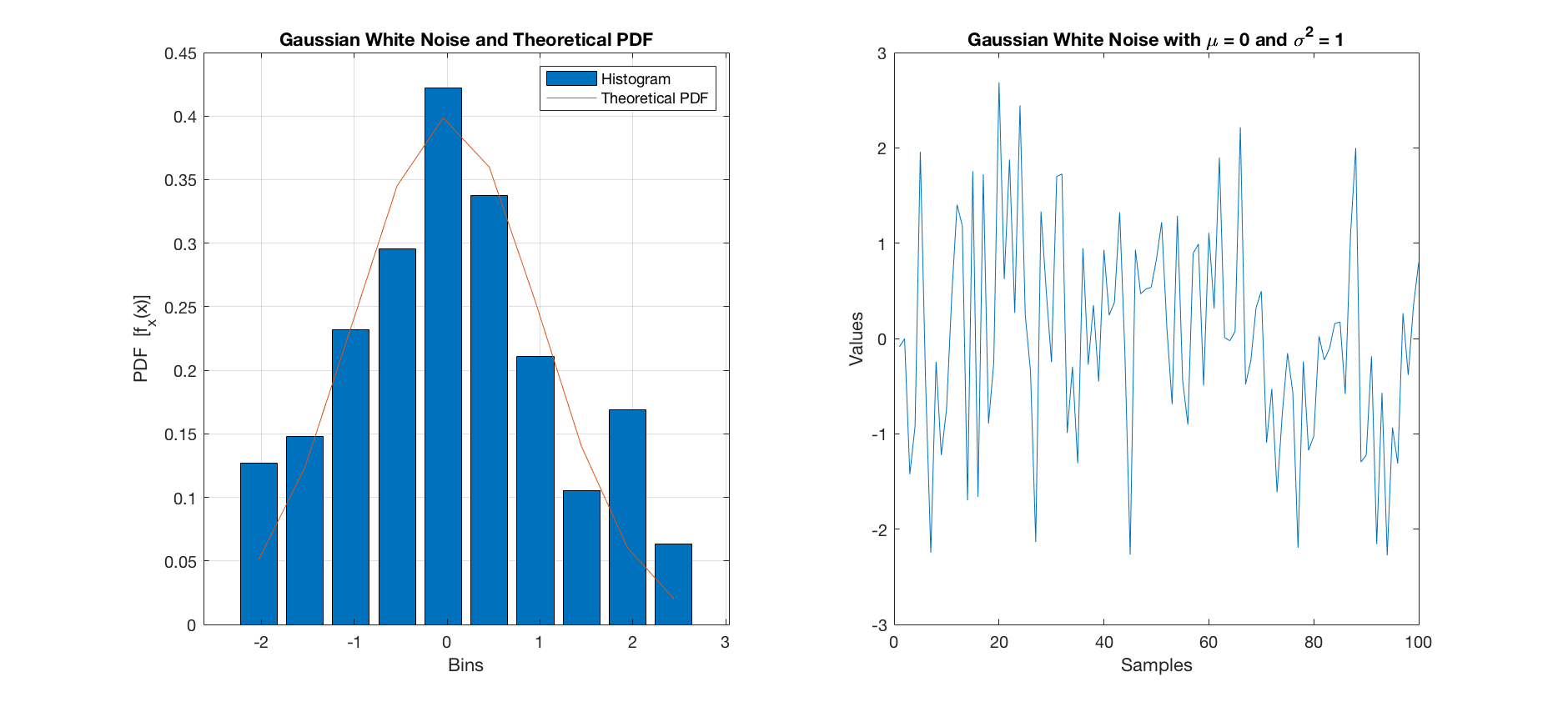}
	\caption{Gaussian White Noise Distribution}
	\label{gwn01}
\end{figure}
\end{center}
As previously shown in \cite{akc4}, the flow along the line of symmetry represents the unperturbed flow; with increasing distance away from this line and closer to the boundary walls, perturbation grows nonlinearly. Such dependence on randomness requires a randomized version of the operator $\mathcal{L}$ that would have stochasticity embedded within:
\begin{equation}
\label{gov18}
\bf{\mathcal{L}}  = \bf{\mathcal{L}} _s + \bf{\mathcal{L}} _u,
\end{equation}
where $\bf{\mathcal{L}} _s$ is the stead state matrix of the system and $\bf{\mathcal{L}} _u$ is the stochastic differential operator. 
For simplicity, we assume ${\bf{\mathcal{L}} _u}$ to be a function of the amplitude ($\epsilon$) and Gaussian white noise ($\bf{X}\sim {\mathcal {N}}(0 ,\sigma ^{2})$), where $\bf{X}$ is a random variable as shown in the figure (\ref{gwn01}).
Also, the random variables are statistically uncorrelated, having a normal white noise with a zero mean and finite variance:  
\begin{subequations}
\begin{align}
\label{gov19}
\bf{\mathcal{L}}  &= \bf{\mathcal{L}} _s + \epsilon \:(\bf{X}\sim {\mathcal {N}}(0 ,\sigma ^{2}))
\\
&= \bf{\mathcal{L}} _s  + \epsilon{ } \:\bf{\xi}(t). \label{gov20}
\end{align}
\end{subequations}
{$\bf{\xi}(t)$} is a diagonal matrix with entries drawn from the matrix
 $x_{ij} \in \{(\bf{X}\sim {\mathcal {N}}(0 ,\sigma ^{2}))$, where  $i = j$. Equation (\ref{gov17}) thus becomes:
\begin{equation}
\label{gov21}
\lambda
\begin{bmatrix}
u_r \\
u_\theta\\
u_z \\
\end{bmatrix}
=
\begin{bmatrix}
\mathcal{L}^*_{rr}          & \mathcal{L}_{r \theta}          & 0  \\
\mathcal{L}_{\theta r} & \mathcal{L}^*_{\theta \theta}  & 0  \\
0                     &                0                          & \mathcal{L}^*_{zz}  
\end{bmatrix}
\begin{bmatrix}
u_r \\
u_\theta\\
u_z 
\end{bmatrix} 
+ 
\begin{bmatrix}
D \\
\frac{i\beta}{r}\\
i\alpha\\
\end{bmatrix}p,
\end{equation}
where
\begin{subequations}
\begin{eqnarray}
\mathcal{L}^*_{rr}  &=& \mathcal{L}_{rr} + \xi_u(t)  \, \label{gov22}\\
\mathcal{L}^*_{\theta \theta} &=& \mathcal{L}_{rr} + \xi_{\theta}(t) \, \quad \& \label{gov23}\\
\mathcal{L}^*_{zz} &=& \mathcal{L}_{rr}  +  \frac{1}{r^2}  + \xi_{z}(t)  \label{gov24}
\end{eqnarray}
\label{govv}
\end{subequations}

The dynamics of the Fourier spectra, defined by the triad $(\xi_u(t), \xi_\theta(t), \xi_z(t)),$ represent the sought after stochastically perturbed equilibrium dynamics.

\section{Numerical Scheme}
In general, studying the Navier-Stokes equations as a dynamical system necessitates the implementation of complex numerical software. This software is divided into two main branches: modeling, that identifies steady states (including uniformly moving waves) and their (in-)stability characteristics, and Direct Numerical Simulation (DNS). In the modeling approach one makes an ansatz, based on the anticipated nature of the flow, and then proceeds to solve a truncated Fourier expanded system. Computing equilibria requires the use of Newton’s method, but since the computation of the required matrices is computationally time consuming due to its size, we rely on iterative Krylov subspace methods. These methods can compute an approximate solution to the Newton equation through repeated integrations on trial solutions, that have commenced based on the initial (educated) guess.

The calculations become possible, because at moderate $Re$, the state space dynamics is low dimensional, making the corresponding matrices sparse. This class of systems are well suited for a class of algorithms called Krylov subspace methods which require computations only of the matrix-vector products. Once the system's numerical solution has been identified, its stability is analyzed via the Arnoldi iteration, which is a Krylov subspace method used for computed the eigenvalues and eigenvectors of a given matrix. Essentially, the algorithm uses stabilized Gram-Schmidt iteration to produce an orthogonal basis for the Krylov subspace. Arnoldi iteration tends to find the most dominant eigenvalues of a matrix first, so it could be terminated, prior to finding all eigenvalues and eigenvectors of a matrix. For the purposes of computing the stability of an equilibrium, this is desirable since it is the least stable eigenvectors of a state that determine the dominant behavior in its neighborhood.

DNS, on the other hand, makes no such assumptions, but is based on first principles with no other mathematical assumptions. As such, DNS is more computationally expensive, but recent advancements in hardware make it a more reasonable approach than before. Since the Navier-Stokes equations do not have an explicit evolution equation for the pressure, it must either be eliminated or solved for in such a way that incompressibility is maintained by the DNS algorithm. Modeling approaches, such as \cite{agb}, can reduce the computational demand, by implementing the conservation law ab initio \cite{Chandra1960}, thus eliminating questions about integrating the Navier-Stokes equations, by choosing the right method for enforcement of the incompressibility constraint. In fact the DNS developed in \cite{agb}, based on the methodology of \cite{Chandra1960}, has been used in the present work, where the computation of the pressure at each time step, is not necessary. This simplifies the calculations and converges to solutions faster.
An accurate numerical computation based on an expansion in terms of Chebyshev polynomials yields an accurate solution for the identification of the required eigenvalues. 

This work thus uses a hybrid Chebyshev spectral collocation method as in \cite{tc12} and solves the eigenvalue problem (\ref{gov11}-\ref{gov14}) in order to retain many of the advantages of the schemes described at the beginning of this section. The choice of this method, over some more technically \enquote{advanced} methods, such as \cite{tc01, tc05}, is guided by the need to emphasize the linearized flow spectrum, that should be reproducible from traditional methods, without any loss of generality. Alternative methods, like Alvaro's Petrov-Galerkin method using basis functions derived from the continuity equation satisfying the boundary conditions \cite{tc01}, or Simon et al's method \cite{tc02} based on Legendre polynomials rather than Chebyshev polynomials in constructing the basis function, do provide faster convergence but, do not improve on resolution which is our key requirement. Our work uses a more direct numerical solution of the linearized Taylor-Couette flow problem simultaneously satisfying the incompressibility condition, thus leading to the matrix solution
\newpage
\begin{equation}
\label{gov25}
\lambda \bf{B} \bf{q} = \bf{A} \bf{q},
\end{equation}
where 
\begin{equation}
\label{gov27}
\bf{q} =  
\begin{bmatrix}
u_r \\
u_\theta\\
u_z \\

\end{bmatrix}, \quad 
\bf{B} =
\begin{bmatrix}
I  & 0 & 0 & 0 \\
0  & I & 0 & 0 \\
0 & 0 & I & 0 \\
0  & 0 & 0 & 0 
\end{bmatrix}, \quad \text{and} \quad
\bf{A} = 
\begin{bmatrix}
\mathcal{L}^*_{rr}          & \mathcal{L}_{r \theta}          & 0   & D\\
\mathcal{L}_{\theta r} & \mathcal{L}^*_{\theta \theta}  & 0    & \frac{i\beta}{r} \\
0                     &                0                          & \mathcal{L}^*_{zz}  &  i\alpha \\
D_+ & \frac{i\beta}{r} & i\alpha & 0
\end{bmatrix},
\end{equation}

leading to
\begin{equation}
\label{gov26}
\lambda  = \bf{B}^{-1}\bf{A}.
\end{equation}

The boundary condition is constrained by:
\begin{equation}
\label{gov28}
\bf{C} \bf{q} = 0 \quad \text{and} \quad
\bf{C}
=
\begin{bmatrix}
u_r|_{r_i = 0} & 0 & 0 & 0  \\
u_r|_{r_0 = 0}  & 0 & 0 & 0 \\
0 & u_{\theta}|_{r_i = 0}  & 0 & 0 \\
0  & u_{\theta}|_{r_0 = 0}& 0 & 0  \\
0 &0 & u_z|_{r_i = 0} & 0 \\
0  & 0& u_z|_{r_o = 0} & 0 
\end{bmatrix}.
\end{equation}

\noindent
Equations (11-14) and  (\ref{gov21}) combined define the system matrix given in Eq. (\ref{gov27}). Equation (\ref{gov28}) is numerically solved by constraining the Jacobian matrix $\bf{A}$ and the partial Identity matrix $\bf{B}$ with the constraining matrix $\bf{C}$, together with the boundary conditions.

In order to compute  the temporal flow variation, sample data are generated numerically from the expression: 
 \begin{equation}
 \label{gov29}
 {\mathbf{\Lambda_{max}}} = \mathcal{F}(Re_i, \epsilon, \eta, N),
 \end{equation}
where ${\mathbf{\Lambda_{\text{max}}}}$ represents the \enquote{maximum eigenvalue} matrix of dimension $\eta_n \times N$. For each $\eta$, there exist $N$ number of $\lambda_{\text{max}}$'s.
The temporal correlation $R$ defining the strength of correlation between the maximum eigenvalues, effectively quantifying the strength of stochastically driven radial correlations, can be computed from Eqns. (\ref{gov26}-\ref{gov28}) as follows:
\begin{equation}
\label{gov30}
R(\tau) =  < [\lambda_{max}(\eta, t + \tau) - \lambda_{max}(\eta, t) ]^2> _{\eta,t}.
\end{equation}
Note, that apart from being the largest in value, $\lambda_{\text{max}}$, is also the most unstable eigenvalue
 \quad and $\tau$ is the spatial separation. The corresponding temporal correlation coefficient $\vartheta$ is defined as
$\vartheta \sim \frac{\ln R }{\ln \tau}$.

\section{Results}
In this paper, the least stable eigenvalues and their corresponding eigenvectors have been considered (in the same spirit of \cite{Balbus1991}). For the purpose of analyzing the role of stochastic perturbations in linearized nonnormal flows, our attention here is on the specific axisymmetric perturbation configuration, where the azimuthal wave number $(\beta)$ is always zero and the axial  wave number $(\alpha )$  is non-zero. These are the most essential perturbation modes that are known to trigger linear unstable behaviour leading to Taylor Vortices (TV) when the inner cylinder incessantly rotates $(Re_i \neq 0)$ and the outer cylinder is kept constant $( Re_o = 0)$. Also, the eigenvalue are either real or a complex conjugate as stated by DiPrima and Hall, which in fact agrees with the principle of stability exchange \cite{tc14}.
\subsection{Axisymmetric Perturbation Without Noise}
\begin{figure}[h!]
	\centering
	\begin{subfigure}[b]{0.45\linewidth}
		\includegraphics[width=\linewidth]{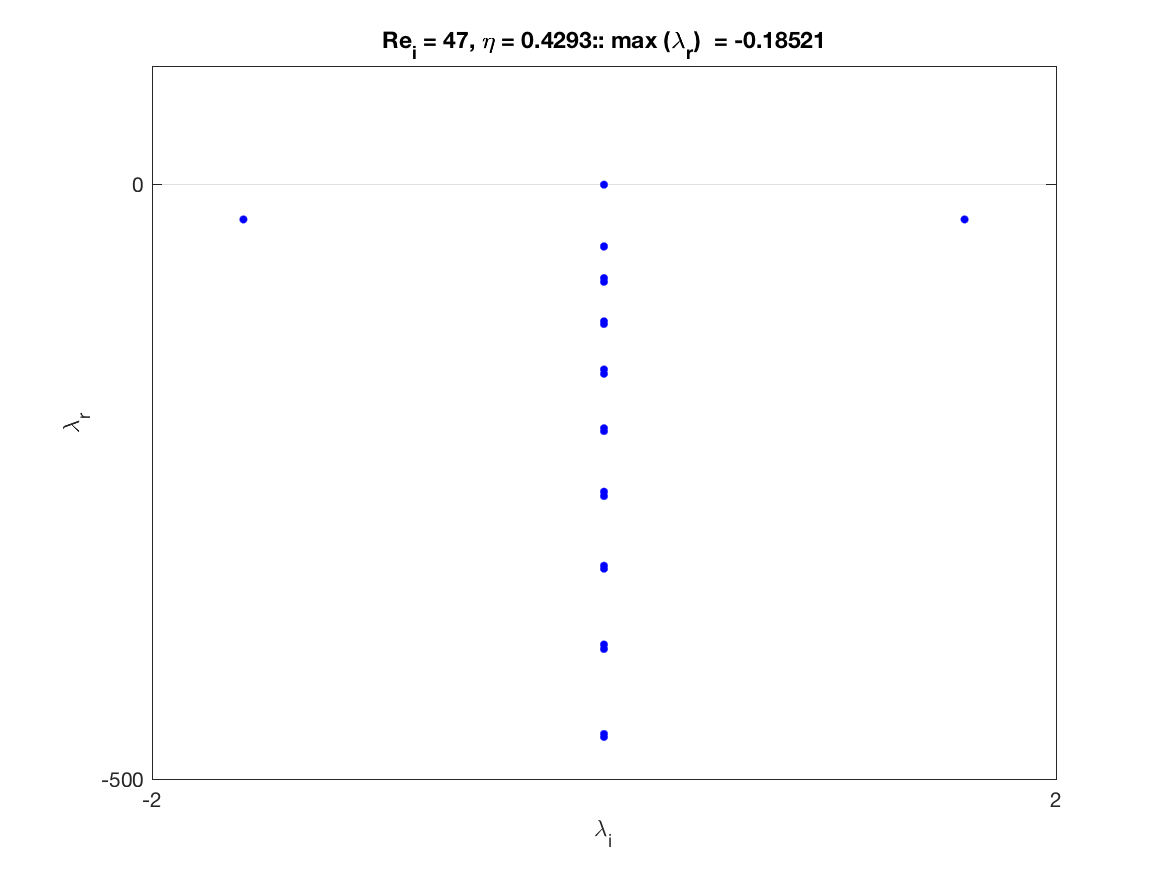}
	  \caption{ }
	\end{subfigure}
	\begin{subfigure}[b]{0.45\linewidth}
		\includegraphics[width=\linewidth]{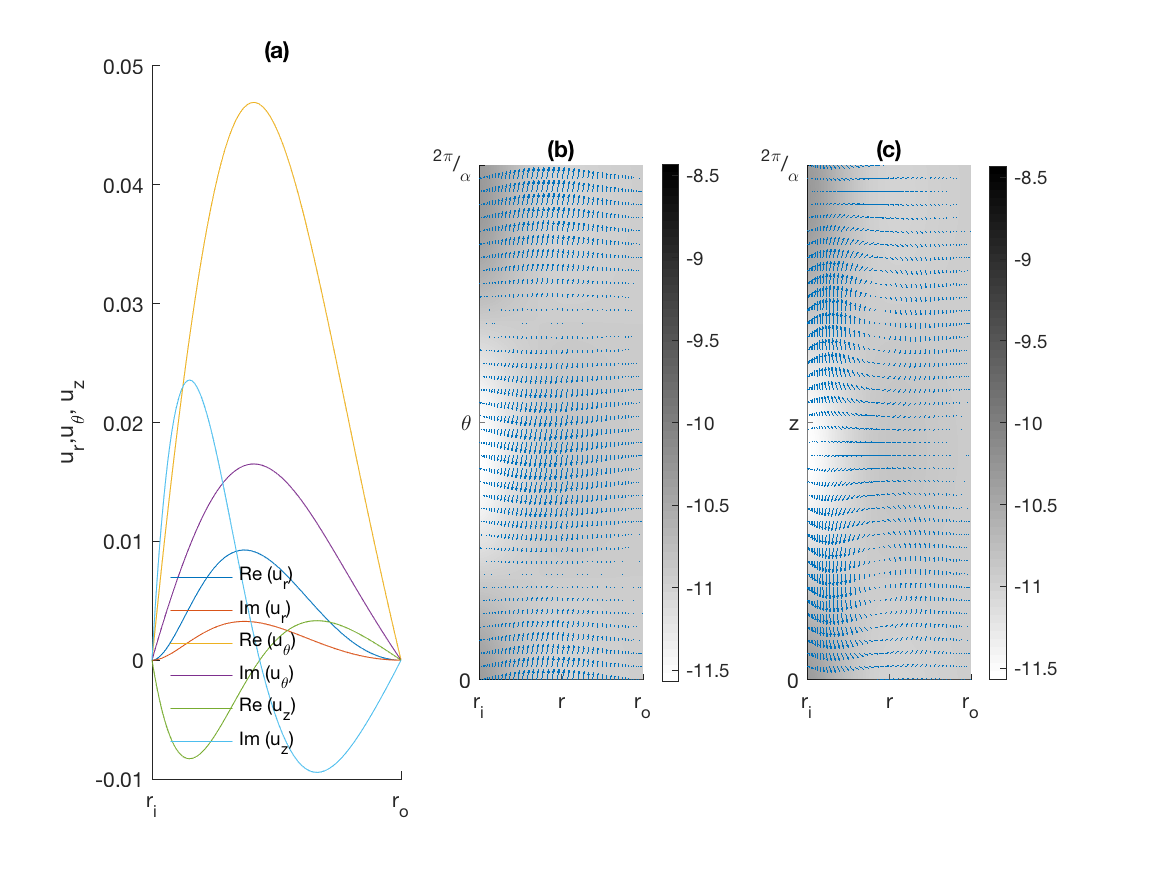}
	    \caption{ }
	\end{subfigure}
	\begin{subfigure}[b]{0.45\linewidth}
		\includegraphics[width=\linewidth]{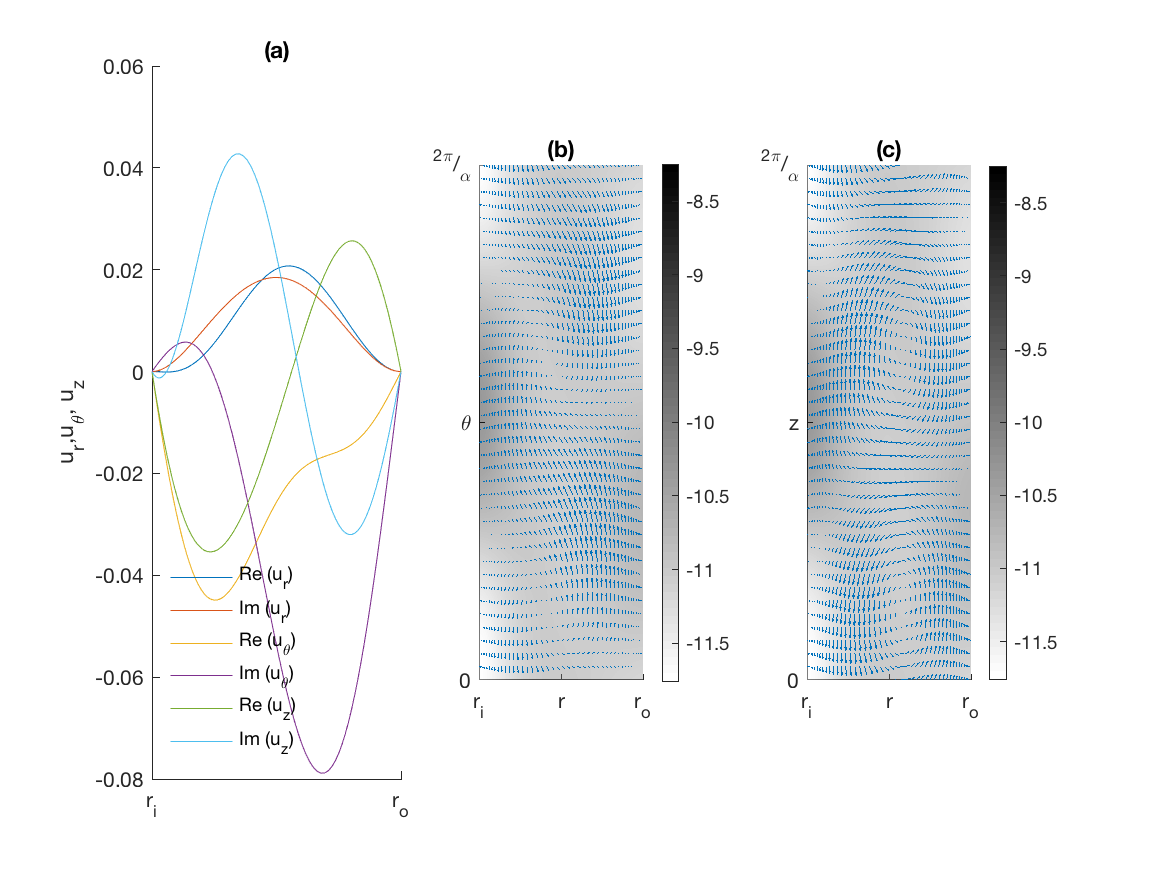}
	    \caption{  }
	\end{subfigure}
	\begin{subfigure}[b]{0.45\linewidth}
		\includegraphics[width=\linewidth]{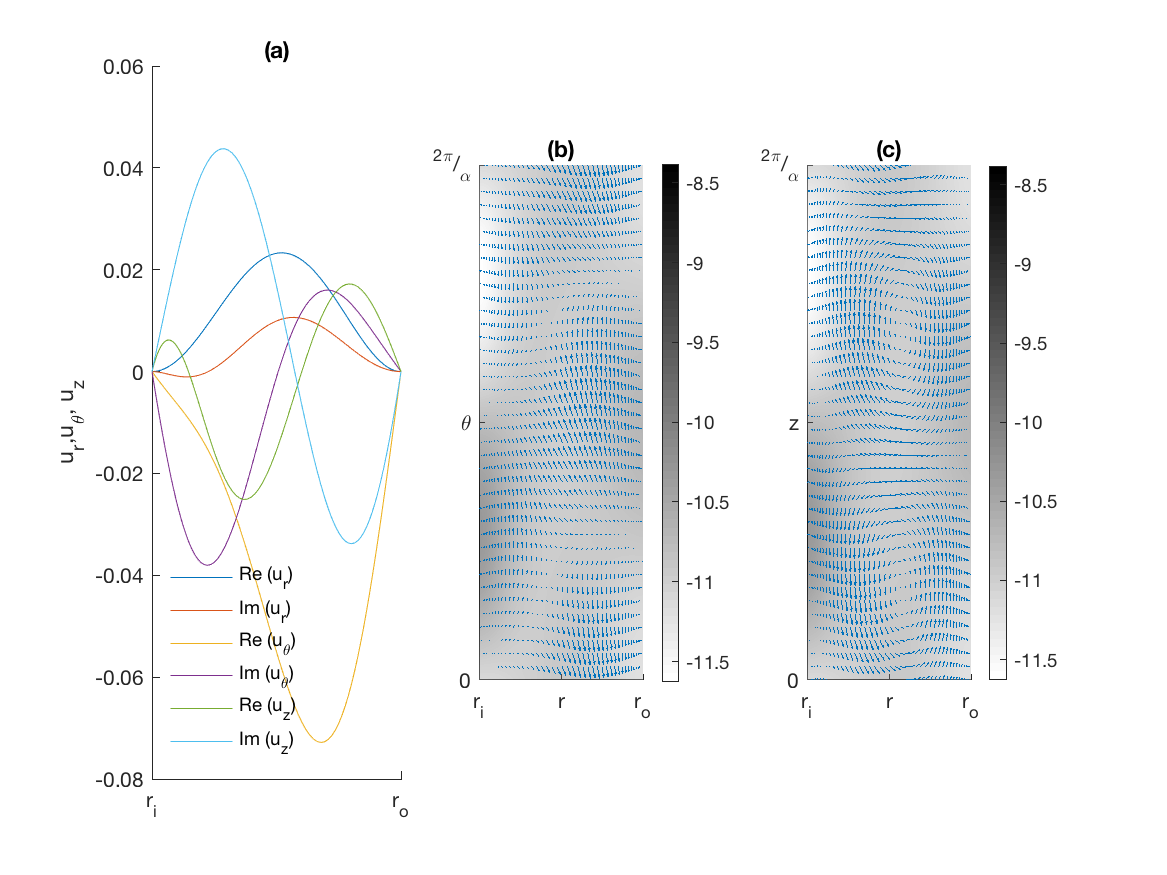}
	    \caption{ }
	\end{subfigure}
	\caption{This figure shows the variation of the eigenvalue, first three eigenmodes, and velocity fields for $Re_i = 47$ and $\eta = 0.4293$. The number of each modes can be clearly seen from (a) to (c) for each plotted velocity field, highlighting the impact of the critical gap in arriving at the instability point.}
	\label{eig01}
\end{figure}

Results from the linearized Navier-Stokes equation for Taylor-Couette flow are found to be in agreement with previous works like \cite{tc02}, \cite{tc04} and more recently \cite{tc03}. 
Our numerical results show the following observation as depicted in figure \ref{fig02}. For $Re_i \le 47$ the flow remains stable with negative real eigenvalue, with no oscillation. But at $Re_i \geq 47$ the flow starts becoming sensitive to altering radius ratios as shown in Fig \ref{eig01}. This is the critical gap of the radius that is slightly larger than that of the gap for which instability is observed for this Reynolds number.

The numerical model is susceptible to the choice of $Re_i$ and $\eta$. We find that linear instability sets in at $Re_i = 47, \:\eta = 0.5224$ and is constrained within the bandwidth $0.15 \geq \eta \leq 0.5224$. At $Re_i = 49$ and $Re_i = 51$, the instability increases to two different radius ratios $\eta = 0.4293$, $0.5224$. The pseudo-first order transition in the phase diagram quantifies the impact of $\eta$ on the flow state where the velocity is restricted to $49<R_i<49$.
\quad Interestingly, proximity to any of these $\eta$ values increases instability for all $Re_i \geq 49$. In the narrow gap regime ({\it i.e.} $\eta = 0.988$), $\forall Re_i < 270$, the flow is stable while for $Re_i \geq 270$, $\forall 0.15 \geq \eta \leq 0.988$, the flow remains unstable. 

\begin{figure}[h!]
	\centering
	\begin{subfigure}[b]{0.45\linewidth}
		\includegraphics[width=\linewidth]{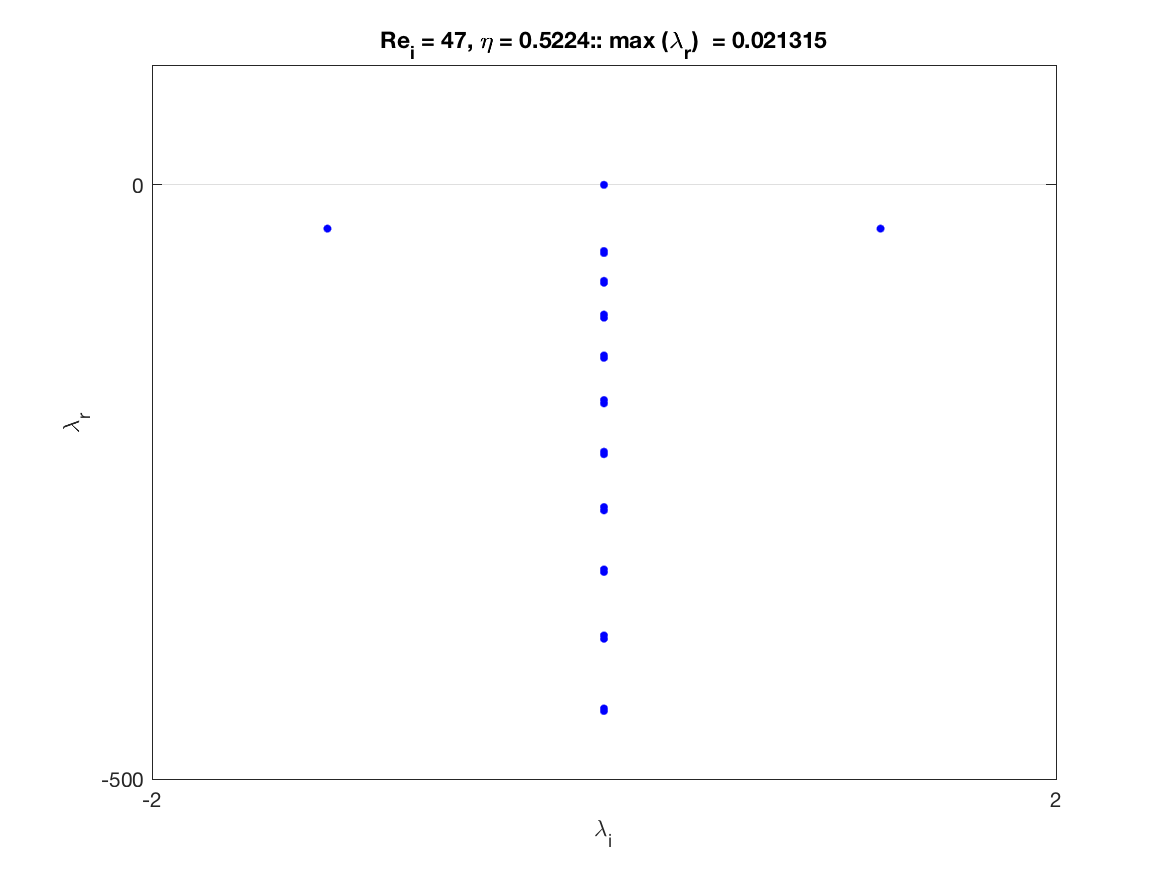}
		\caption{ }
	\end{subfigure}
	\begin{subfigure}[b]{0.45\linewidth}
		\includegraphics[width=\linewidth]{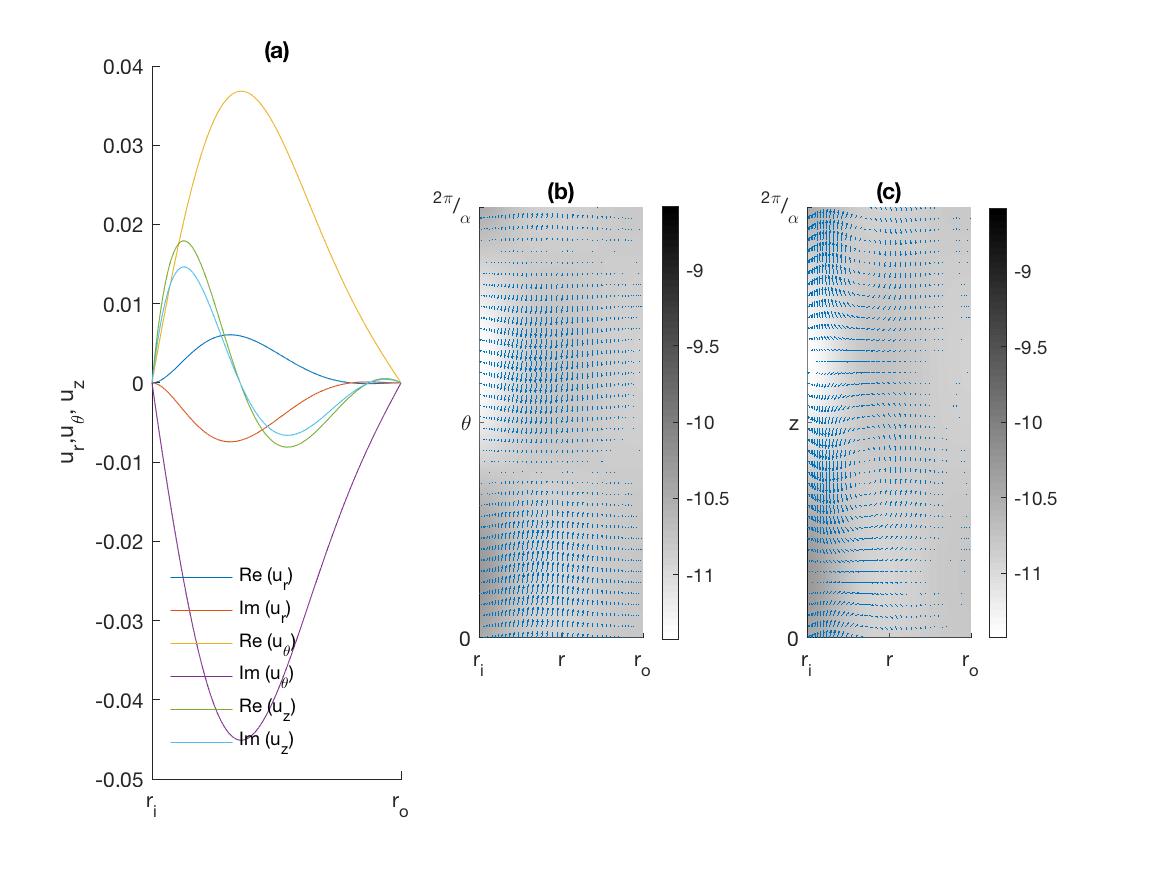}
		\caption{ }
	\end{subfigure}
	\begin{subfigure}[b]{0.45\linewidth}
		\includegraphics[width=\linewidth]{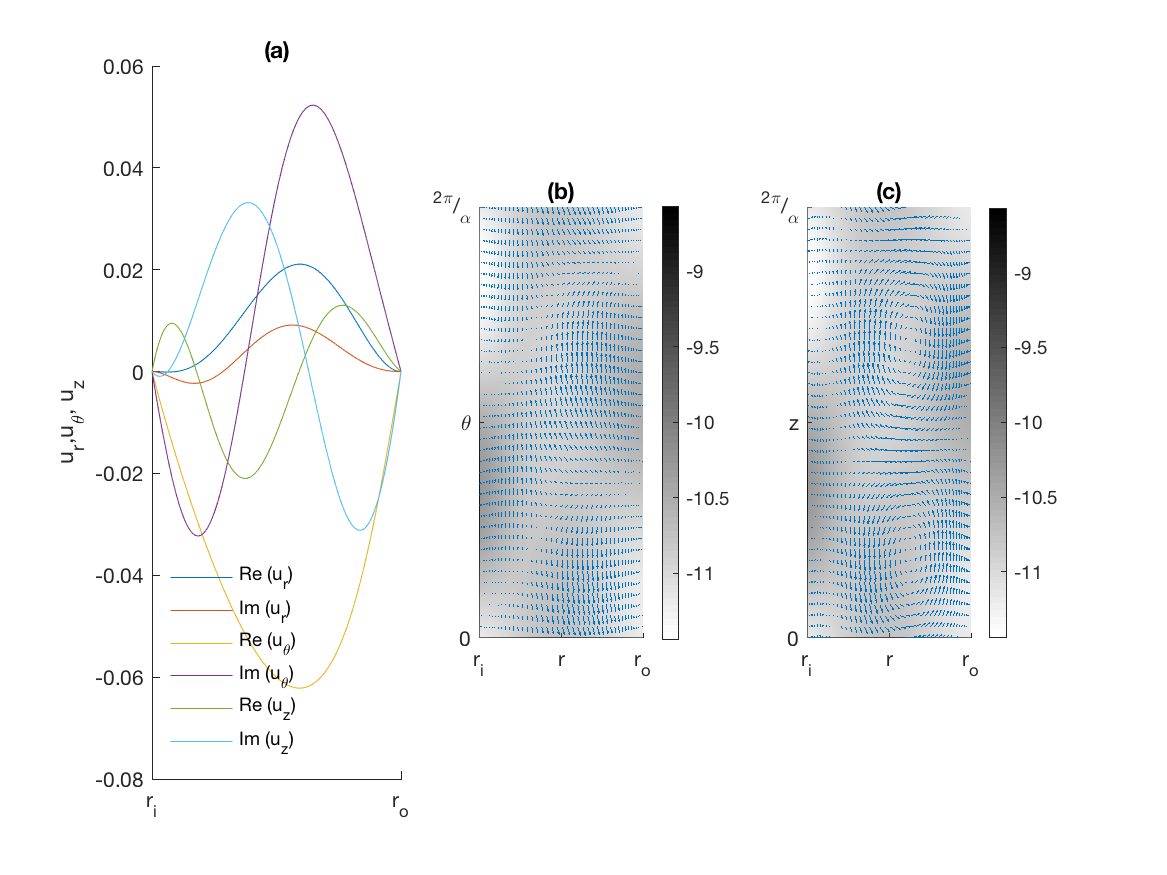}
		\caption{  }
	\end{subfigure}
	\begin{subfigure}[b]{0.45\linewidth}
		\includegraphics[width=\linewidth]{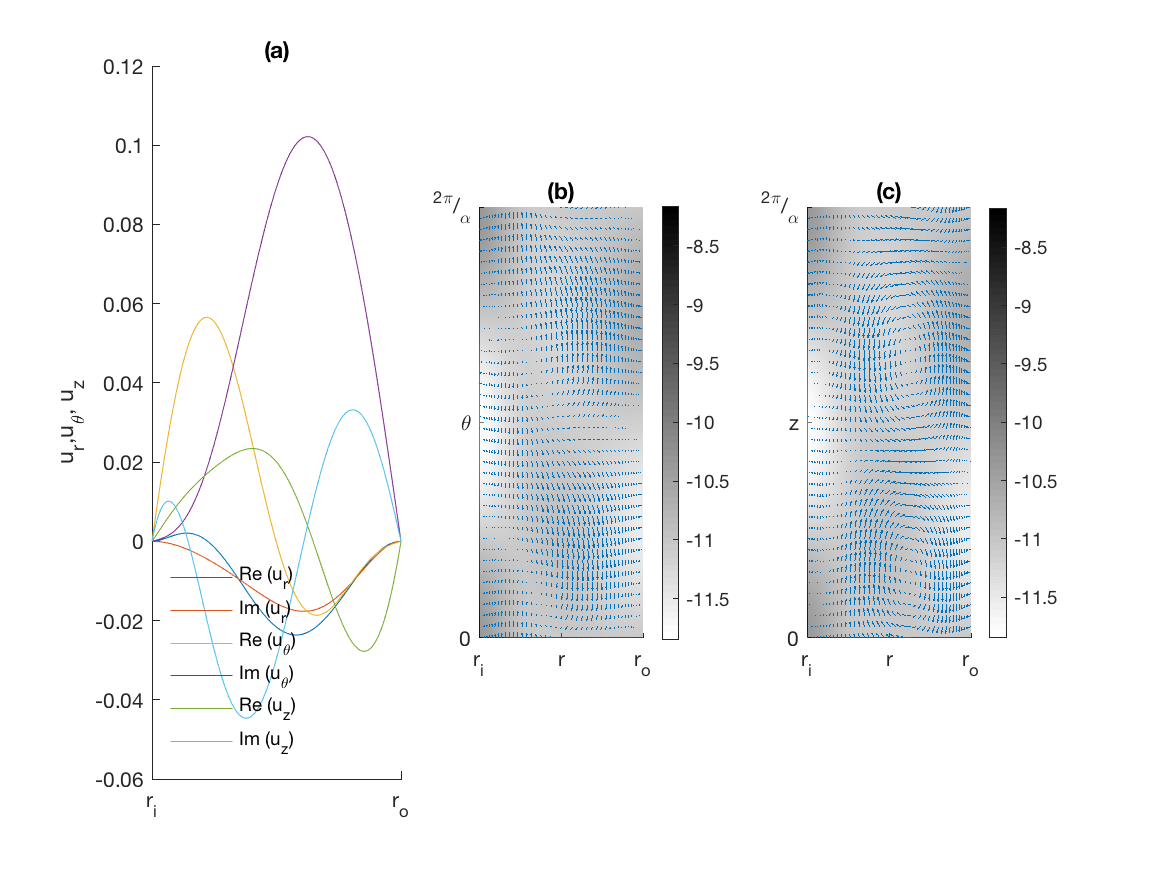}
		\caption{ }
	\end{subfigure}
	\caption{Unstable modes at $Re_i = 47$ and $\eta = 0.5224$.}
	\label{eig02}
\end{figure}
\begin{figure}
	\centering
	\includegraphics[width=0.7\linewidth]{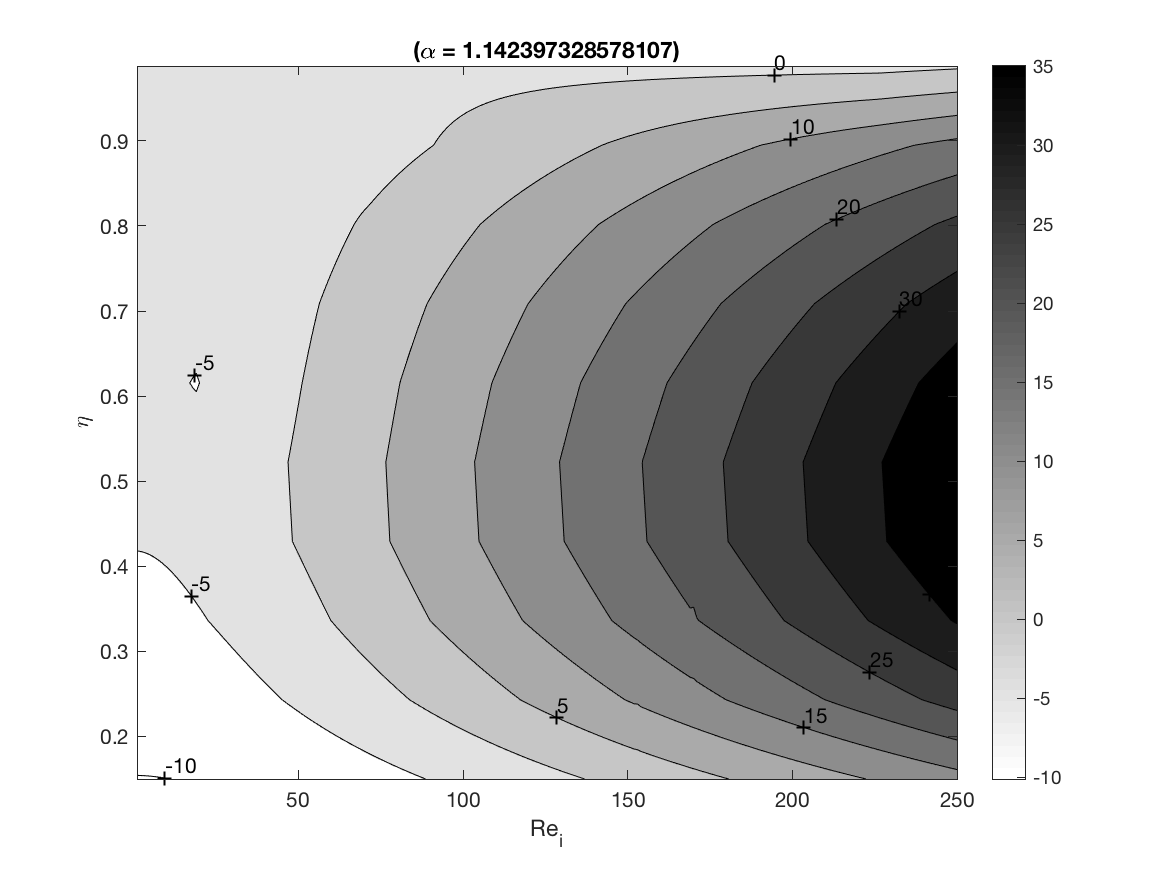}
	\caption{A contour plot showing stable and unstable ranges of $0.15 \geq \eta \leq 0.988$ and $1 \geq Re_i \leq 250$ of the flow. The line $\eta=0$ seperates the stable eigenvalues from unstable ones. The white and light gray colors are the stable region, while the dark gray to black shades represent the unstable eigenvalues. The color bar clearly shows the separations of the flow pattern.}
	\label{fig02}
\end{figure}

\begin{figure}[h!]
	\centering
		\begin{subfigure}[b]{0.45\linewidth}
		\includegraphics[width=\linewidth]{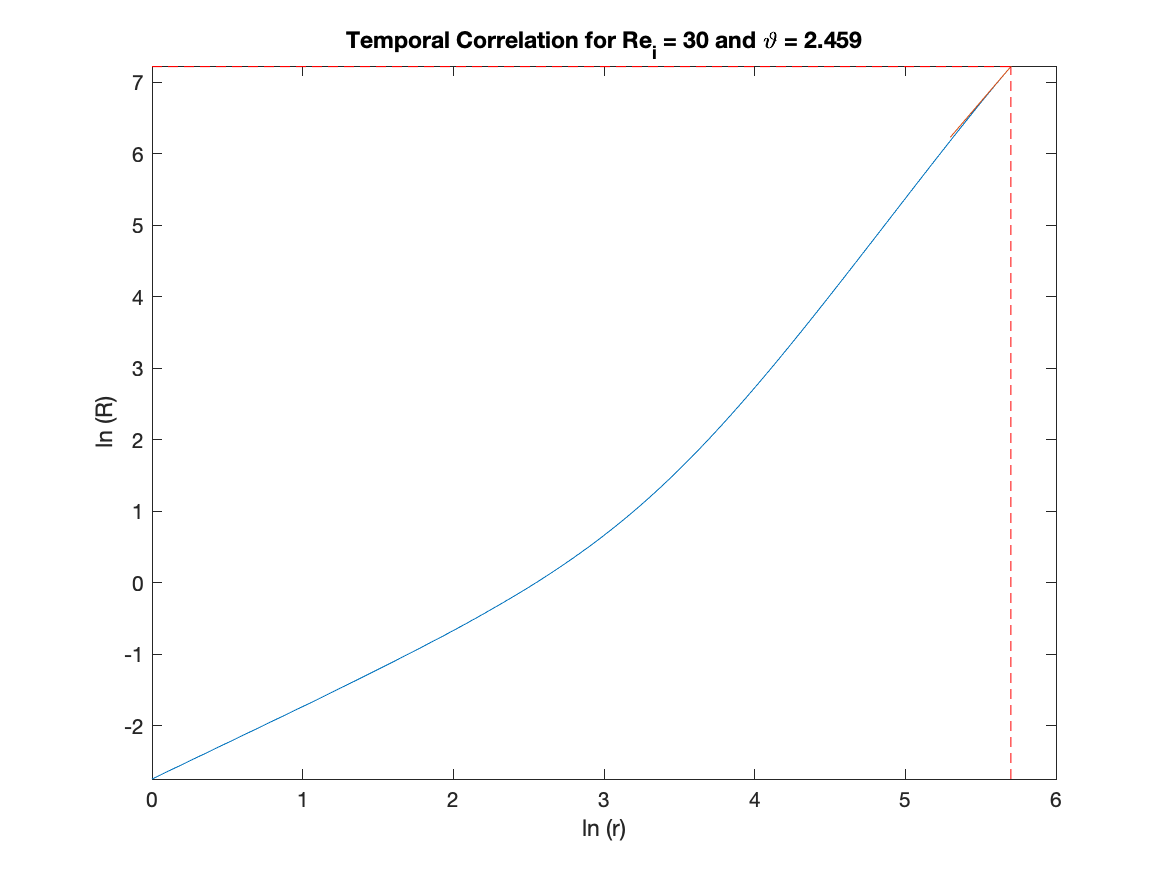}
	\end{subfigure}
	\begin{subfigure}[b]{0.45\linewidth}
		\includegraphics[width=\linewidth]{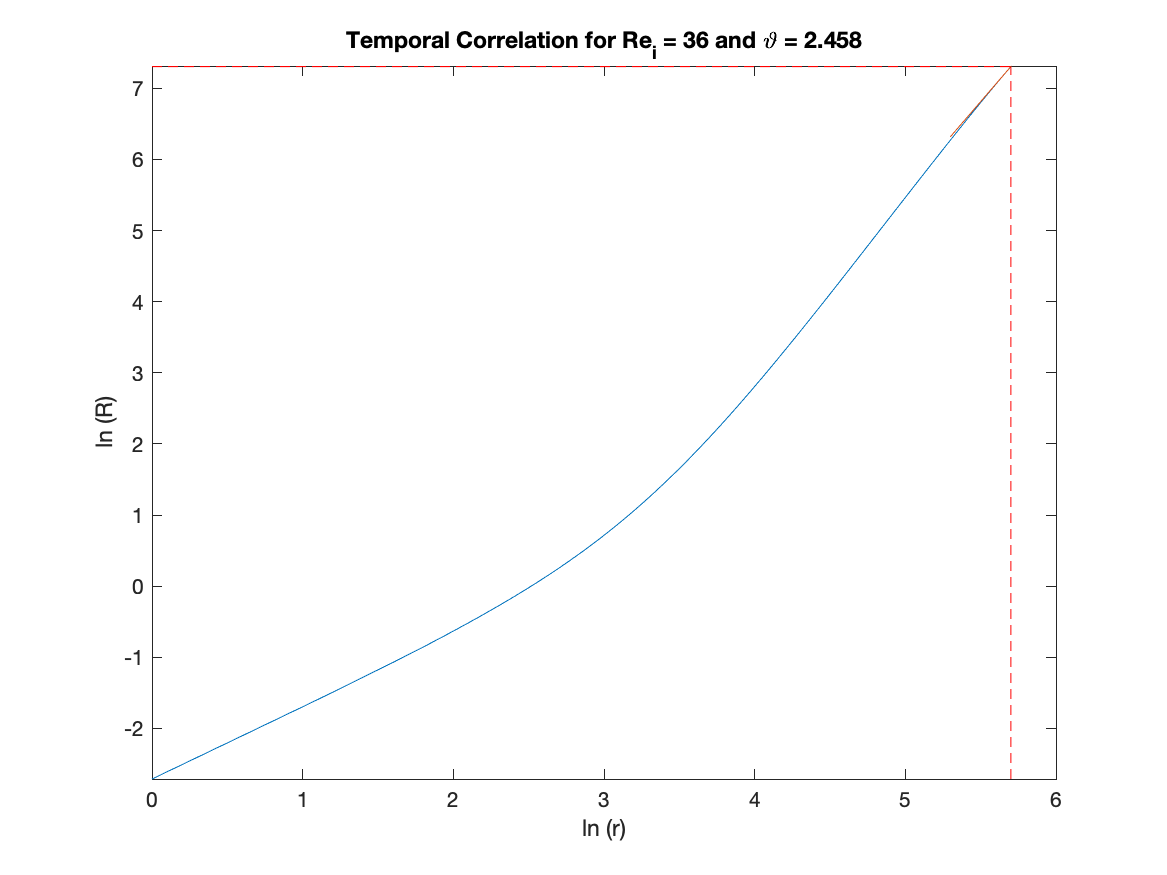}
	\end{subfigure}
	\begin{subfigure}[b]{0.45\linewidth}
		\includegraphics[width=\linewidth]{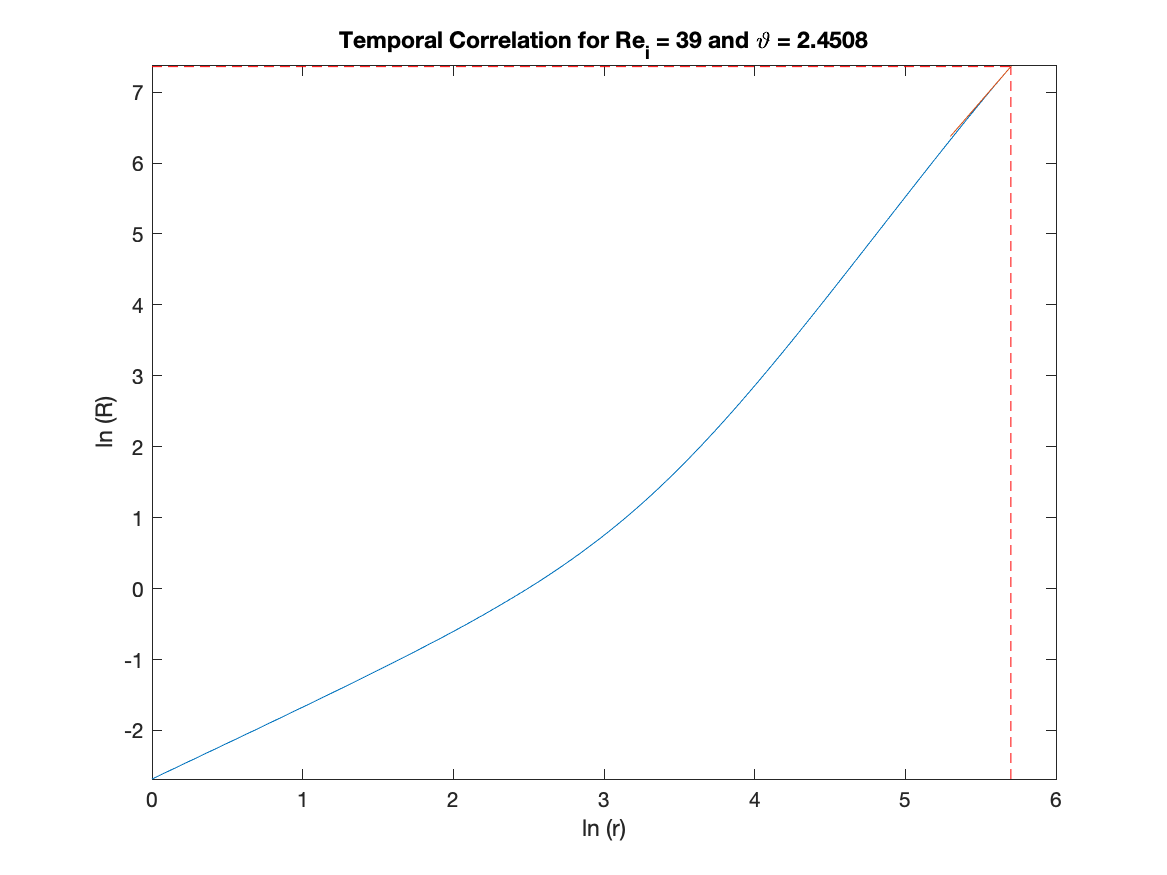}
	\end{subfigure}
	\begin{subfigure}[b]{0.45\linewidth}
		\includegraphics[width=\linewidth]{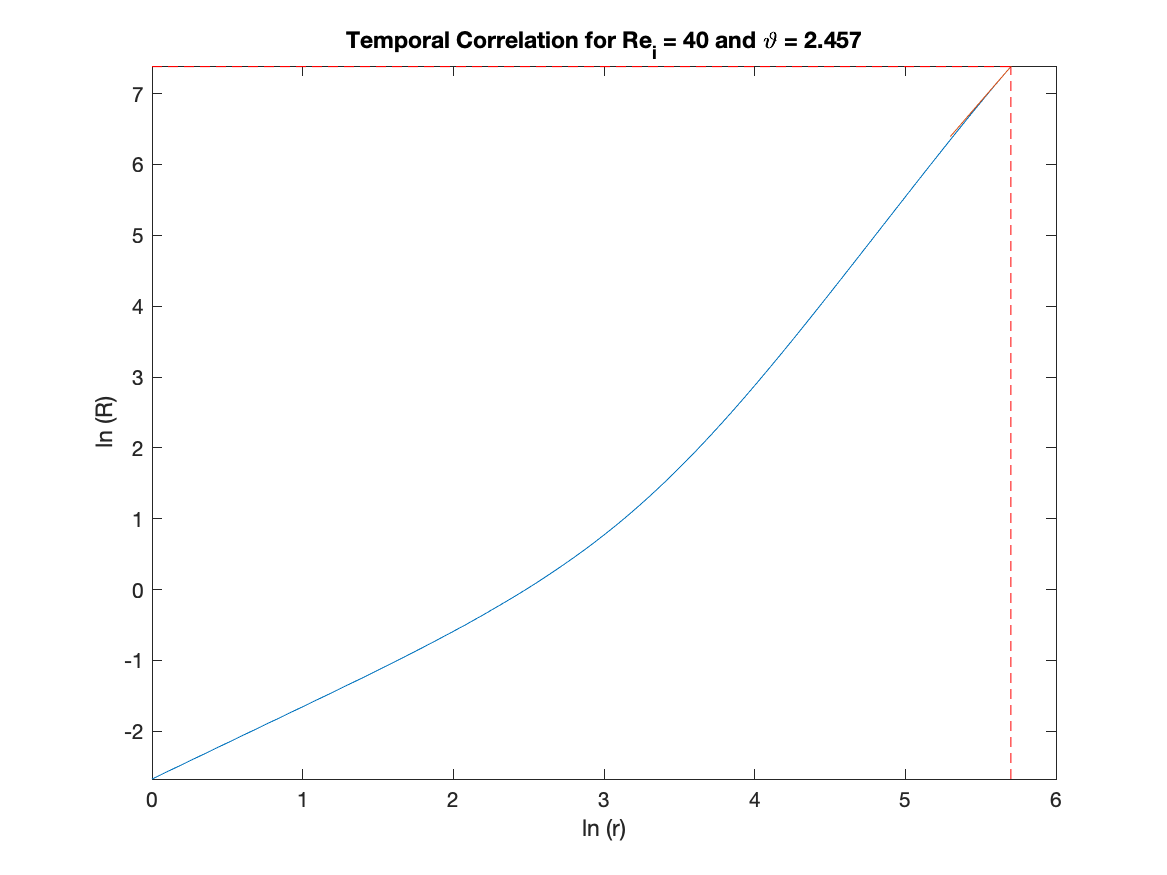}
	\end{subfigure}
	\begin{subfigure}[b]{0.45\linewidth}
		\includegraphics[width=\linewidth]{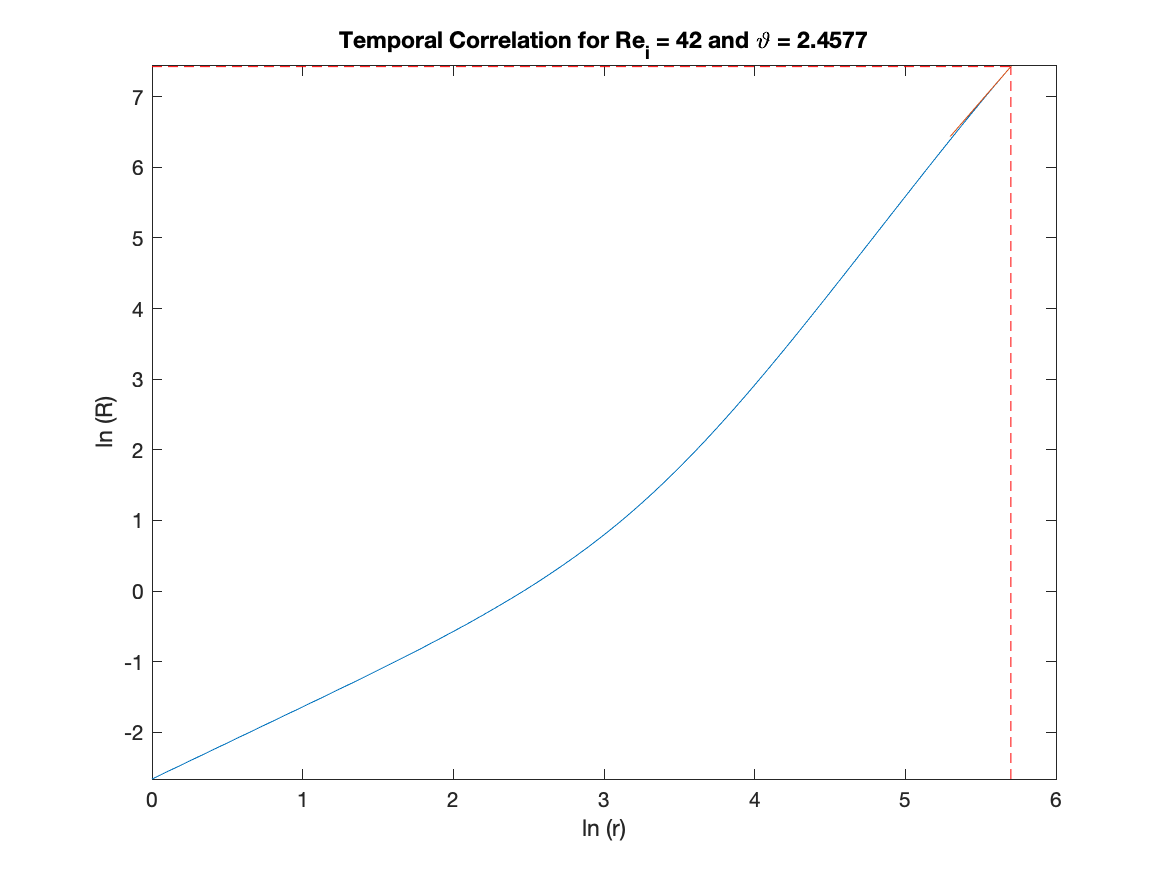}
	\end{subfigure}
	\begin{subfigure}[b]{0.45\linewidth}
		\includegraphics[width=\linewidth]{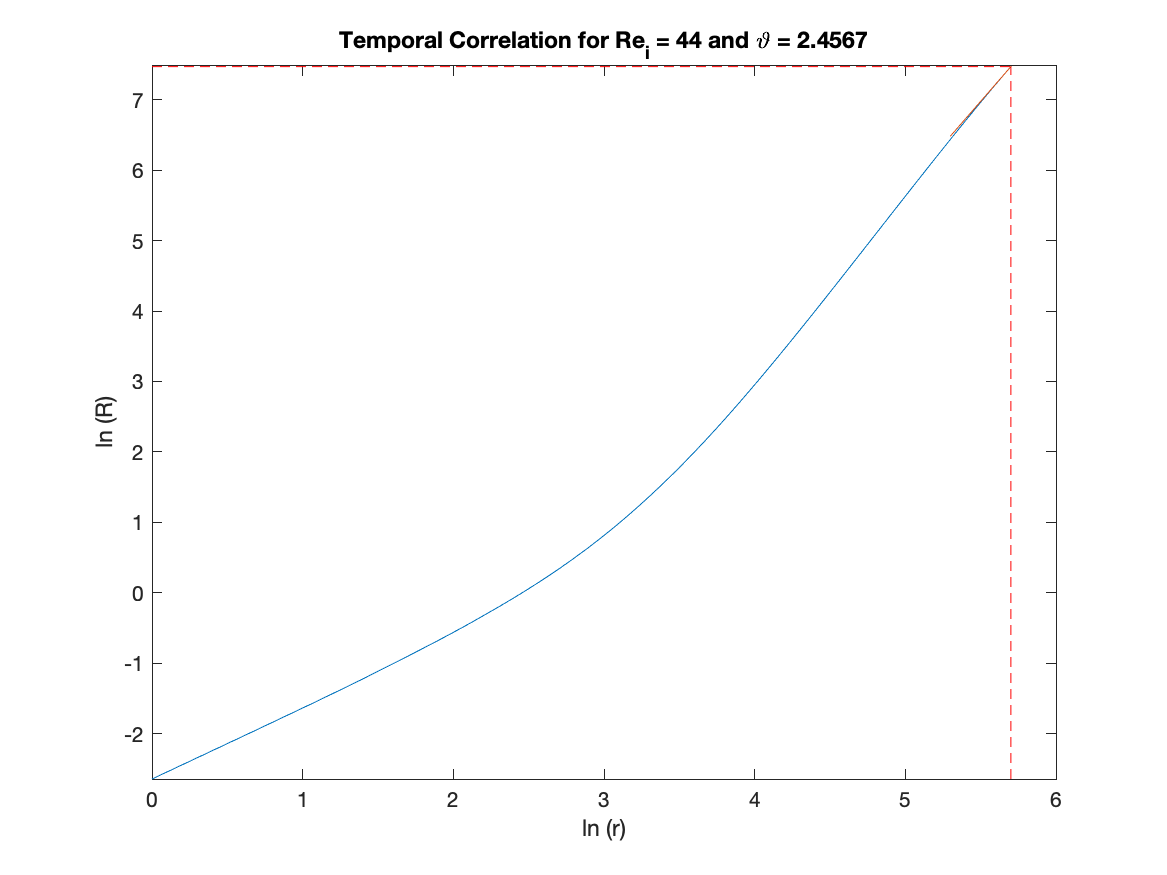}
	\end{subfigure}
	\caption{Max Case: Plots showing the temporal correlation function for various values of $Re_i$ and for a fixed $\epsilon$}.
	\label{sc01}
\end{figure}

\begin{figure} [h!]
	\centering
	\begin{subfigure}[b]{0.8\linewidth}
		\includegraphics[width=\linewidth]{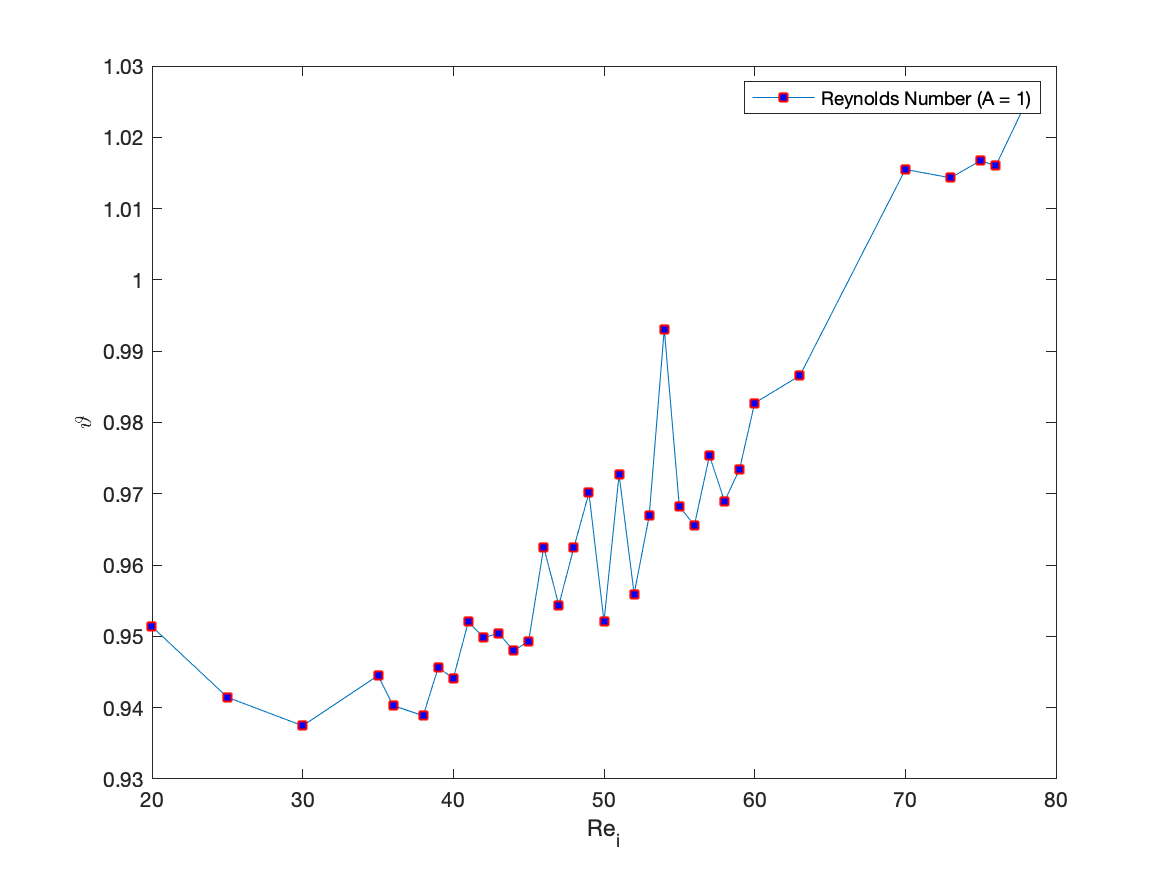}
		\caption{ }
	\end{subfigure}
	\begin{subfigure}[b]{0.8\linewidth}
		\includegraphics[width=\linewidth]{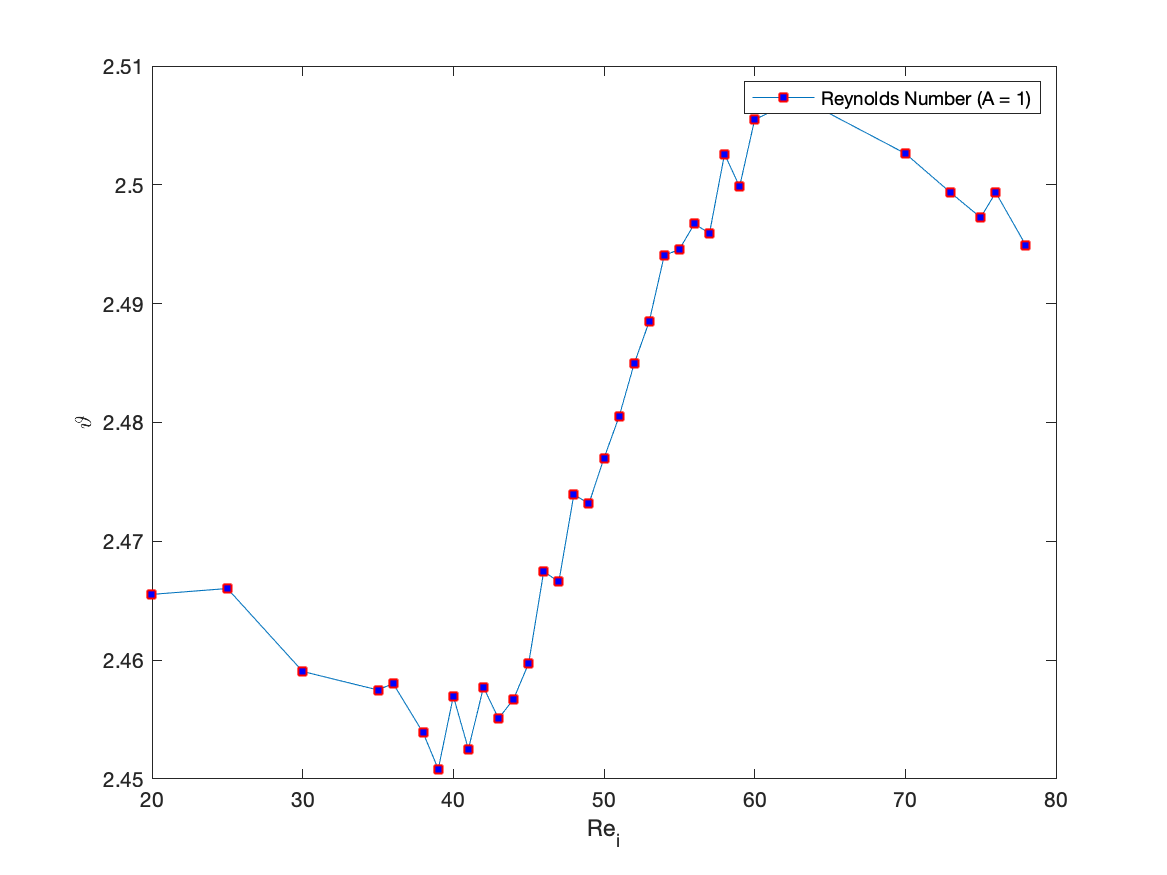}
		\caption{ }
	\end{subfigure}
	\caption{The maximum stability of Taylor-Couette flow is obtained at critical values  
		(a) For $\vartheta_{\text{min}}$-values:  $\vartheta_c \sim 1.0$ and $Re_{i_c} \sim 39$;
	  (b) For $\vartheta_{\text{max}}$-values:  $\vartheta_c \sim 2.5$ and $Re_{i_c} \sim 39$.
        }
	\label{sc02}
\end{figure}

\begin{figure}[h!]
	\centering
	\begin{subfigure}[b]{0.45\linewidth}
		\includegraphics[width=\linewidth]{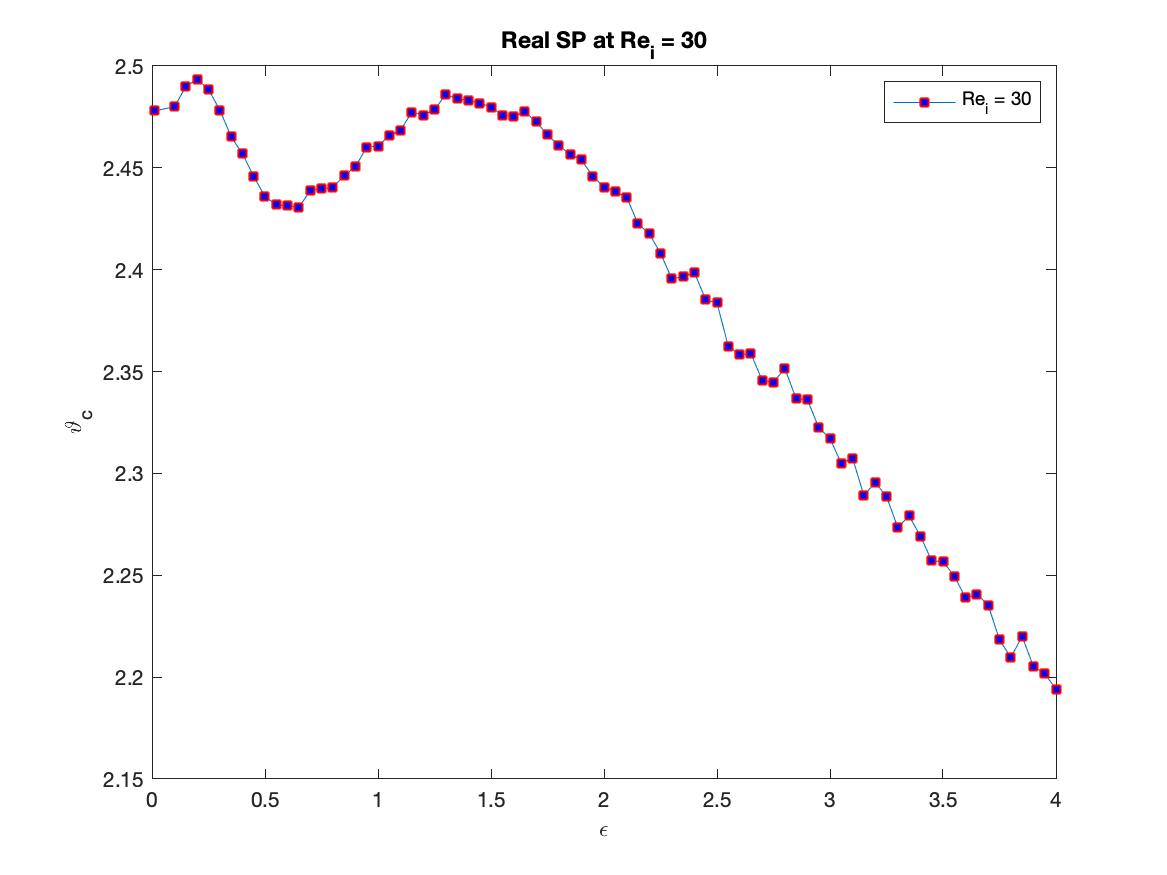}
	\end{subfigure}
	\begin{subfigure}[b]{0.45\linewidth}
		\includegraphics[width=\linewidth]{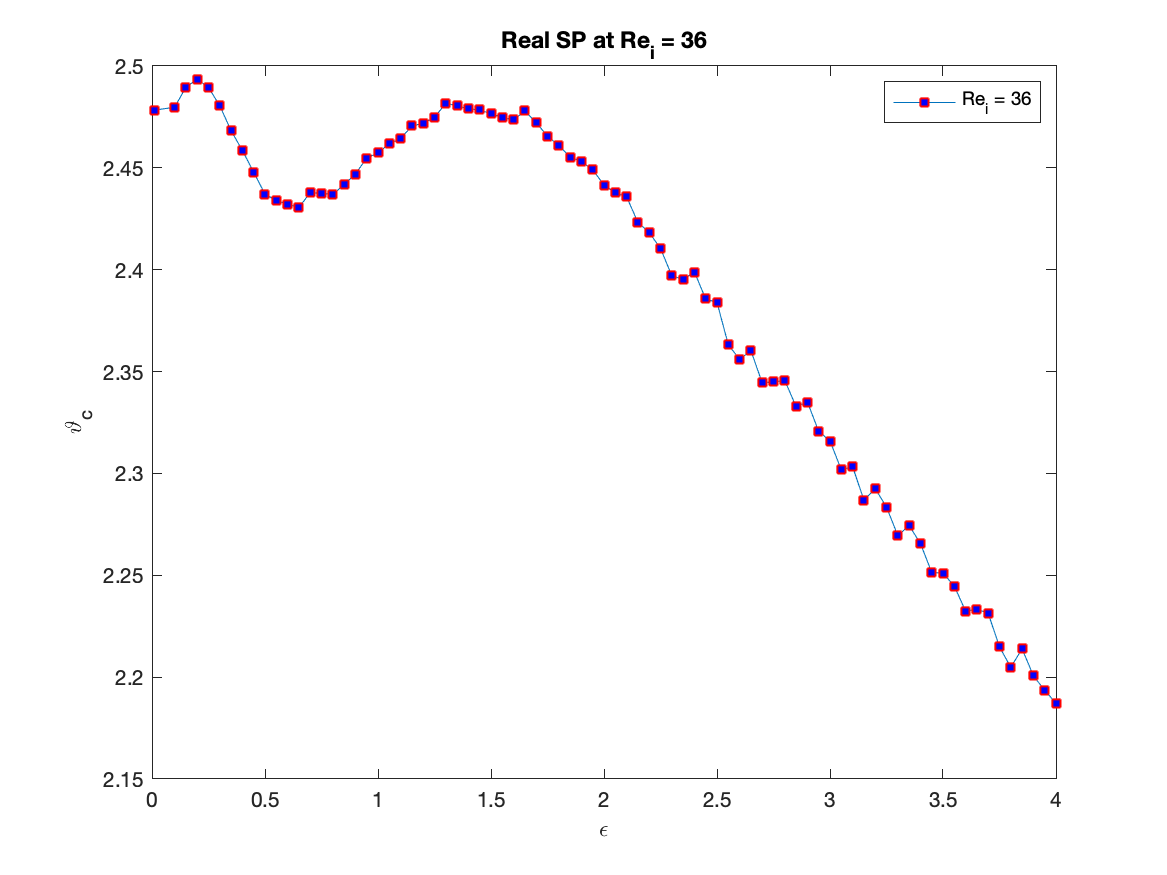}
	\end{subfigure}
	\begin{subfigure}[b]{0.45\linewidth}
		\includegraphics[width=\linewidth]{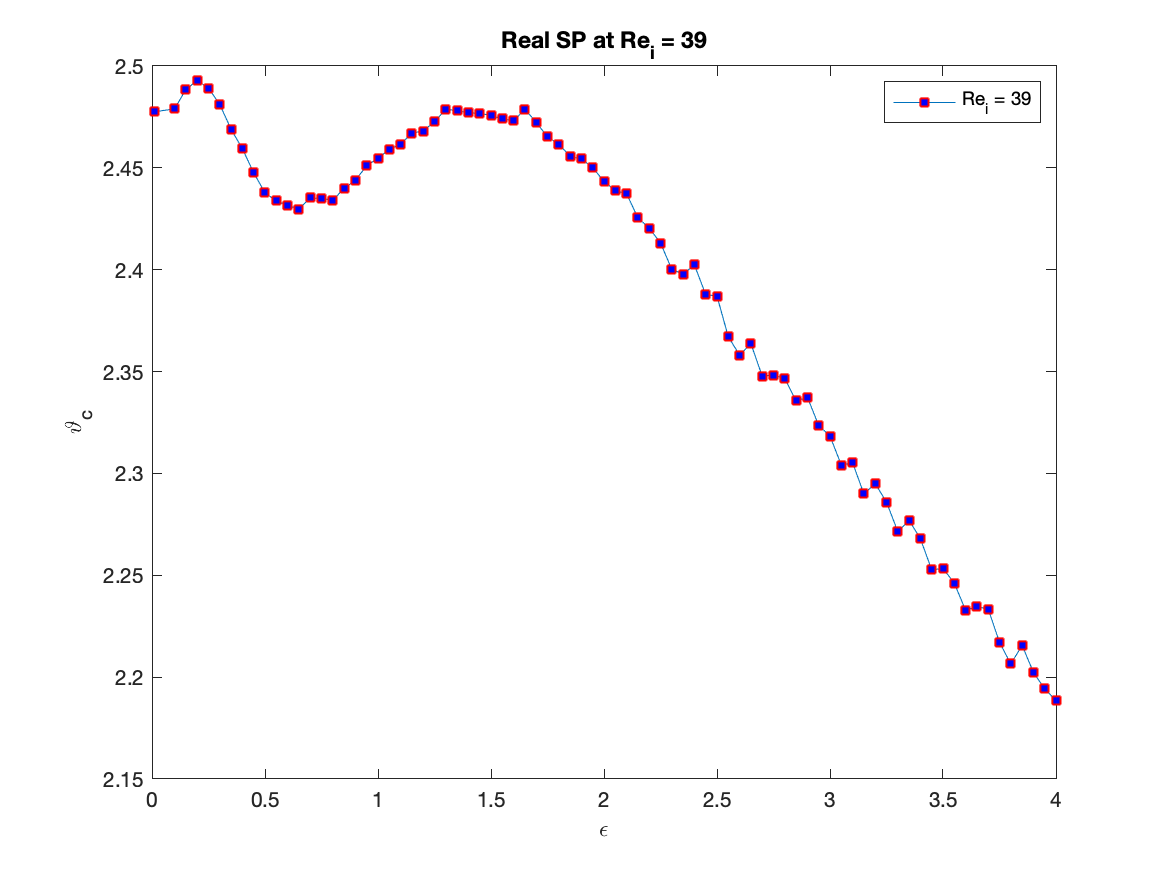}
	\end{subfigure}
	\caption{Plots showing the temporal correlation function for some values of $Re_i$ for a fixed $\epsilon$, showing a local maximum at $\epsilon \sim 0.75$ and a local minimum at  $\epsilon \sim 0.6$}. 
	\label{sc03a}
\end{figure}
\begin{figure}[h!]
	\centering
	\begin{subfigure}[b]{0.45\linewidth}
		\includegraphics[width=\linewidth]{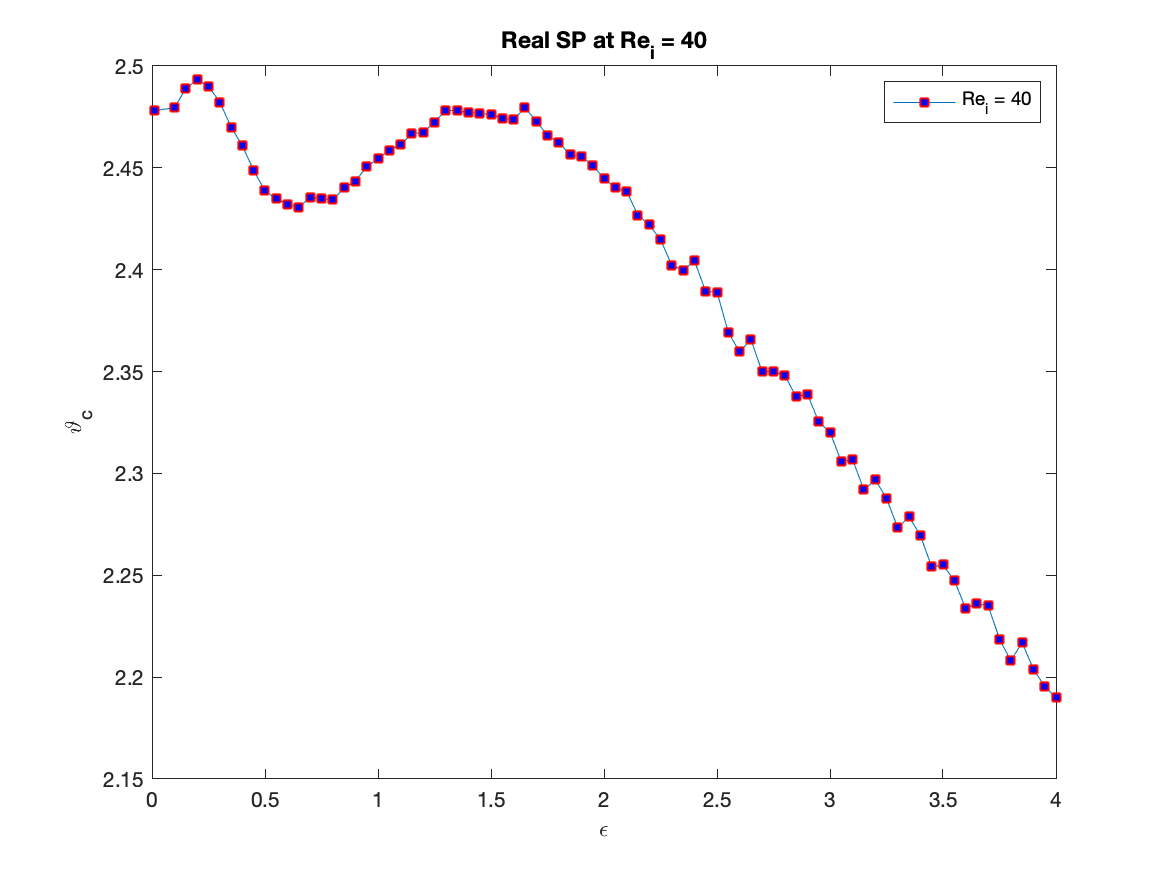}
	\end{subfigure}
	\begin{subfigure}[b]{0.45\linewidth}
		\includegraphics[width=\linewidth]{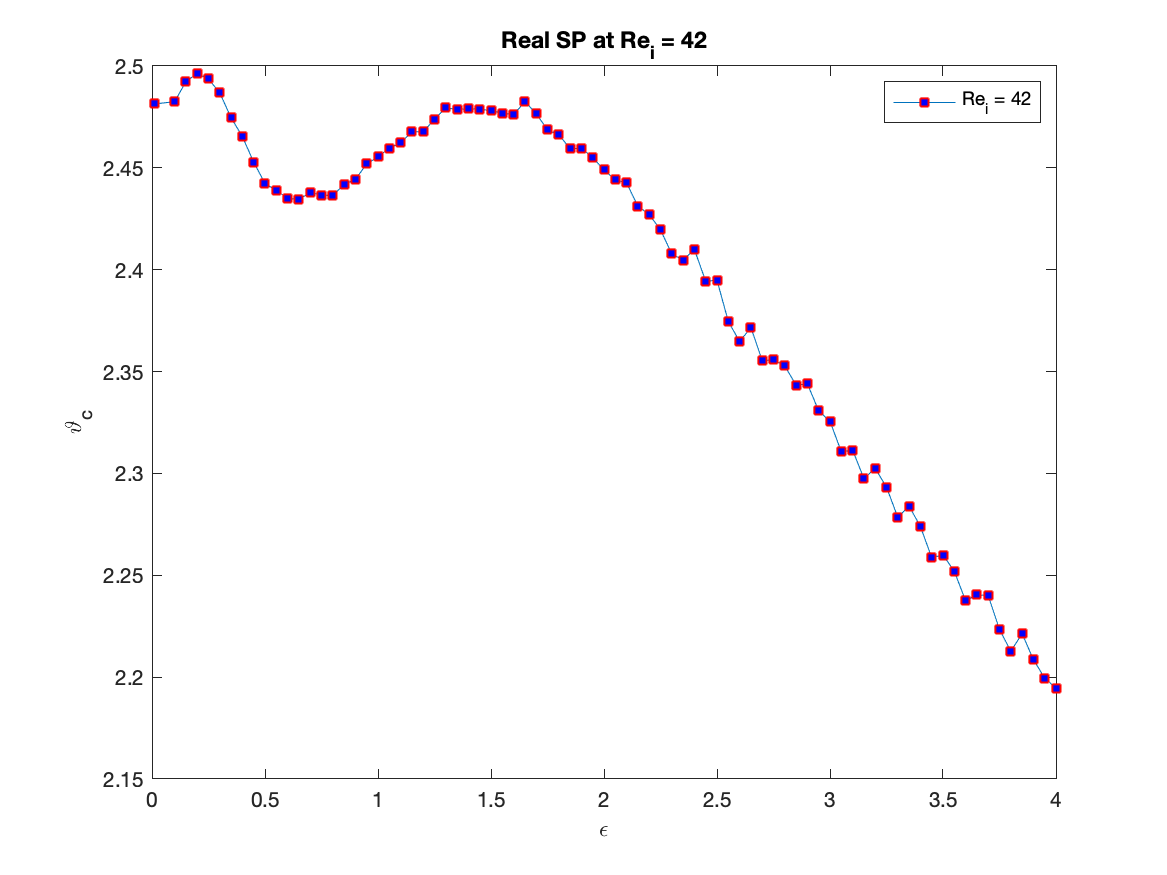}
	\end{subfigure}
	\begin{subfigure}[b]{0.45\linewidth}
		\includegraphics[width=\linewidth]{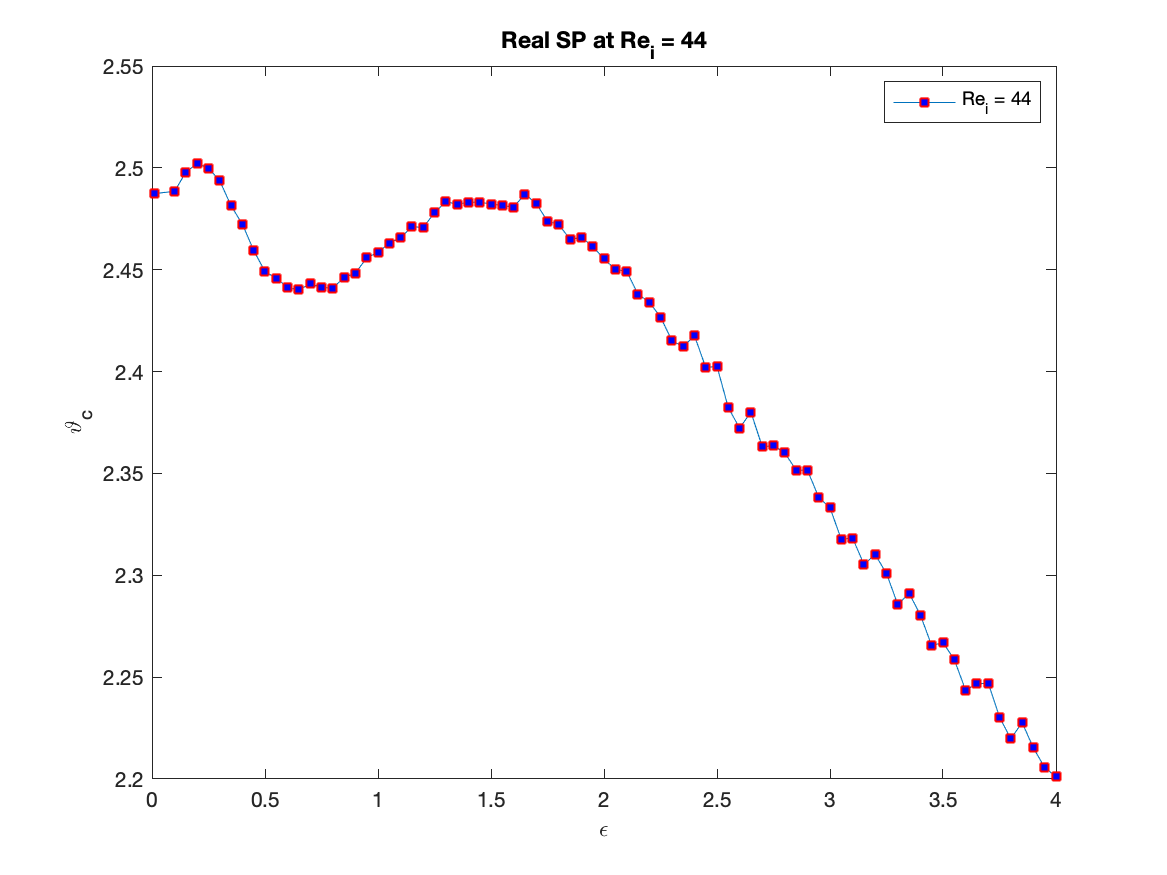}
	\end{subfigure}
	\caption{Plots showing the temporal correlation function for some values of $Re_i$ for a fixed $\epsilon$, showing a local maximum at $\epsilon \sim 0.75$ and a local minimum at  $\epsilon \sim 0.6$}. 
	\label{sc03b}
\end{figure}
\begin{figure}[h!]
	\centering
	\begin{subfigure}[b]{0.45\linewidth}
		\includegraphics[width=\linewidth]{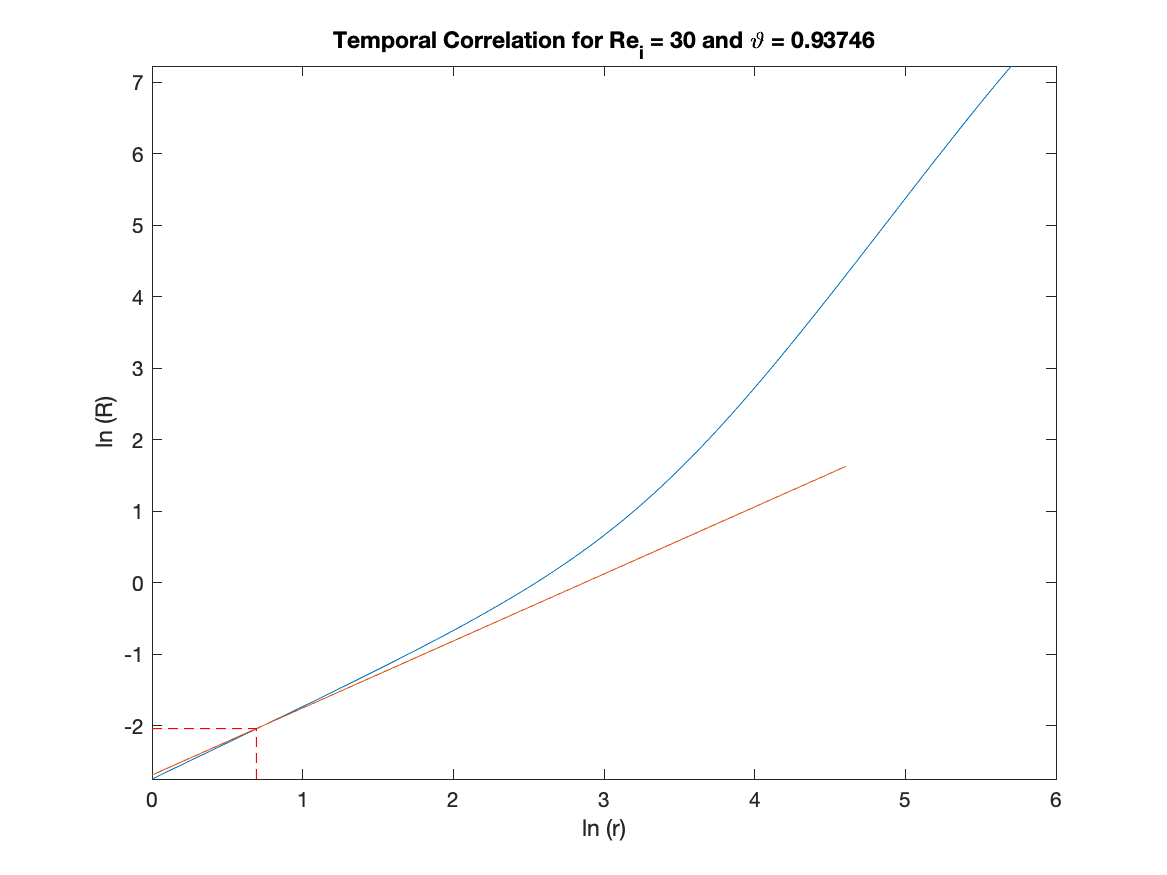}
	\end{subfigure}
	\begin{subfigure}[b]{0.45\linewidth}
		\includegraphics[width=\linewidth]{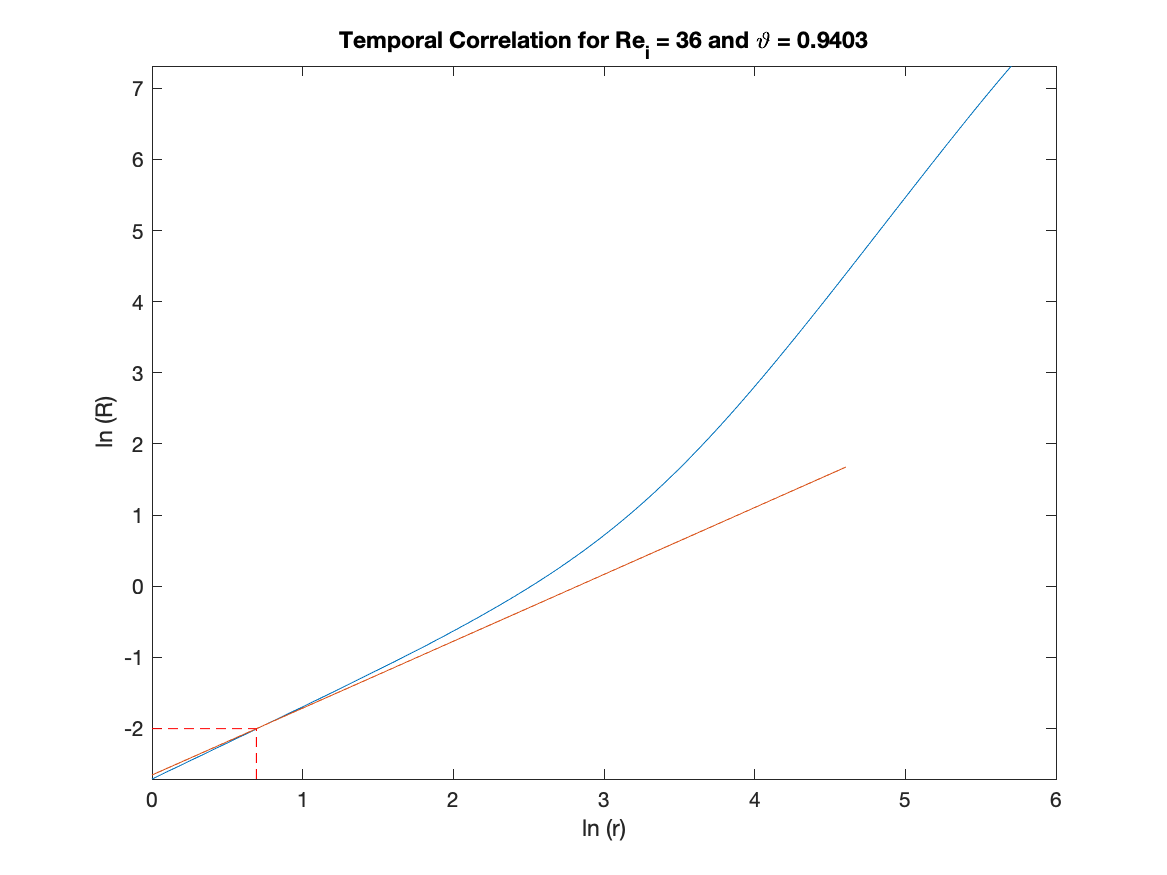}
	\end{subfigure}
	\caption{\enquote{Min} Case: representative plots showing the temporal correlation function for various values of $Re_i$ for a fixed $\epsilon$}
	\label{sc04}
\end{figure}

\subsection{Axisymmetric Perturbation with noise}
The stochastically forced axisymmetric perturbation leads to a maximum for the slope at $\vartheta_{\text{max}} > 1$, while the minimum is represented by $\vartheta_{\text{min}} < 1$.
$R$ scales unboundedly as $\tau^{+\alpha}$ for certain values of the noise amplitude $\epsilon$ and Reynolds number $Re_i$. By keeping $\epsilon$ fixed, $Re_i$ is continuously changed to generate different $\vartheta$ values. These $\vartheta$'s are plotted against the corresponding $Re_i$'s as shown in figure(\ref{sc01}) for the  $\vartheta_{\text{max}}$ and figure (\ref{sc04}) for $\vartheta_{\text{min}}$. 
We choose $\vartheta_{\text{max}}$ as depicted in Figure (\ref{sc01}). This gives the most stable Taylor-Couette flow conformation for $\epsilon = 1$, as shown in Figure (\ref{sc02}). In addition, from the contour plot shown in Figure (\ref{fig02}), the most stable $Re_{i_c} < 50$. From Figure (\ref{sc02}), we choose the $Re_i$'s in the neighbourhood of $Re_{i_c}$. Then for each $Re_i$, we compute the corresponding $\vartheta$ with different values of $\epsilon$, as shown in Figure \ref{sc02} for $\vartheta_{\text{max}}$-values and Figure (\ref{sc05}) for $\vartheta_{\text{min}}$-values.

\section{Conclusion}
This article presents the first of its kind results for stochastically forced eigenvalue dynamics depicting a comprehensive evaluation of the pattern transition at very low Reynolds numbers of the Taylor-Couette flow. We examined the axisymmetric perturbation since its configuration is known to set the transition from laminar to turbulence flow (i.e. to Taylor Vortexes). A subtle pattern formation is found and shows some form of marathon (or a relay of energy) in the transition as the gap increases or reduces depending on $Re_i$. But this pattern decays into an unstable flow, when $Re_i \geq 47$ within certain gap width, then later for all $\eta \in [0.15:0.8948]$ for $Re_i \geq 137$  as shown in the figure (\ref{fig02}). 

The DNS of Navier-Stokes equations at small amplitude (linear) stochastic forcing further revealed the peculiarities of the statistically stationary fluctuating field of a laminar plane Couette flow. Our work showed that due to the non-normal (nonexponential) growth of the hydrodynamic perturbations, their finite statistically defined stationary level is given by the non-normality of the nonequilibrium flow system. Here we found that the anisotropy of the fluctuating velocity field increases with the shear rate. We note that the streamwise constant fluctuations were the main influences on the flow field. In turn, this led to velocity fluctuations that are twice as much as the channel width, that is important for the axisymmetric correlations studied in this work. We also identified nonzero cross-correlations of velocity with streamwise-spanwise components. But we did not find any evidence of the spanwise reflection symmetry breaking reported in \cite{Tahia06}. One may also conclude the existence of some structural regularities in the fluctuating background that can be considered as a seed of some recently observed nonlinear stationary (coherent) structures \cite{agb}.

Stochastically, the axisymmetric flow experiences maximum stability at critical $\vartheta_c$ and $Re_{i_c}$. That is evident from the remarkable similarity of plots in  figures (\ref{sc03a} and \ref{sc03b}). This implies that the linear stability defined at $\Re_{i_c} \sim 39$ (Fig. \ref{sc02}) profile remains largely unaffected by an increasing noise amplitude as shown in panels (a) and (b) of Fig.  \ref{sc02}, respectively indicated by $\epsilon=1.5$). This is not unexpected as the $\vartheta_c$ value is the most stable value for that ($\epsilon, Re_i$) combination as shown in figure (\ref{sc02}b). In addition, from the plots in  figures (\ref{sc03a} \& \ref{sc03b}), the local maximum occurs at $\epsilon \sim 0.75$ ($\vartheta_c \sim 1.5$) and a local minimum at $\epsilon \sim 0.6$ ($\vartheta_c \sim 0.5$). Together, they identify a \enquote{window} of linear stability in a stochastically driven Taylor-Couette flow within which temporal correlation are noise driven, known as {\it constructive coherence}. Figure (\ref{sc02}a) also suggests that the linear stable region for $\Re_{i_c} \sim 39$, even at values of  $\vartheta < \vartheta_c$ when the  $\vartheta_{\text{min}}$-values are considered. 

The stochastically perturbed model defines a two-phased system, one of which is characterized by the temporal correlation $R \sim \tau$ that is independent of the values of $\epsilon$ and $Re_i$, while the other phase will reach $\vartheta_c \rightarrow 0$ for $\epsilon \rightarrow \infty$. But since $\vartheta_c \sim 1.5$, there is a noise independent phase for which $Re_i$ is neutral as indicated by the plots in figure (\ref{sc05}). Hence the former is the so-called \enquote{min-phase} and the later is the \enquote{max-phase}. Such stability points have a key implication in the overall context of an inherently unstable (deterministic) Navier-Stokes model with non-normal instability. The Afshordi et al \cite{bani} and Avila et al \cite{tc02} articles explained non-normal unstable growth in Orr-Sommerfeld-Squire (OSS) deterministic flow models while their stochastic counterparts in the form of the Chattopadhyay et al models \cite{akc1,akc2,akc3} analyzed both magnetic and non-magnetic accretion flow instability using stochastic OSS models, both involving a key convention term that had major contributions in the results. The present article is the first of its kind to analyze the complementary stability aspect of these studies by showing how a non-conserved stochastically forced Navier-Stokes model without any convective nonlinearity could still converge to stable fixed points, with the implication of stochasticity embodied in a shifted stability point. Another key aspect that came out of this study is the suggestive \enquote{universality} of stochastic forcing, as evident from Figures \ref{sc03a} and \ref{sc03b}. Also against a range of instability values, indicated by the plots in Figure \ref{sc05}, the $\vartheta$-gradient values consistently  converge to 0.944, another universality feature. Such surprising stability, both against stochastic forcing and nonlinearity strengths, are clearly suggestive of underlying {\it stochastic coherence}, a case for future study.

\cleardoublepage
\bibliographystyle{IEEEtran}
\bibliography{tcbibtex.bib}
\newpage
\appendix
\section{Temporal Correlations: Varying Nonlinearity}

\begin{figure}[h!]
	\centering
	\begin{subfigure}[b]{0.45\linewidth}
		\includegraphics[width=\linewidth]{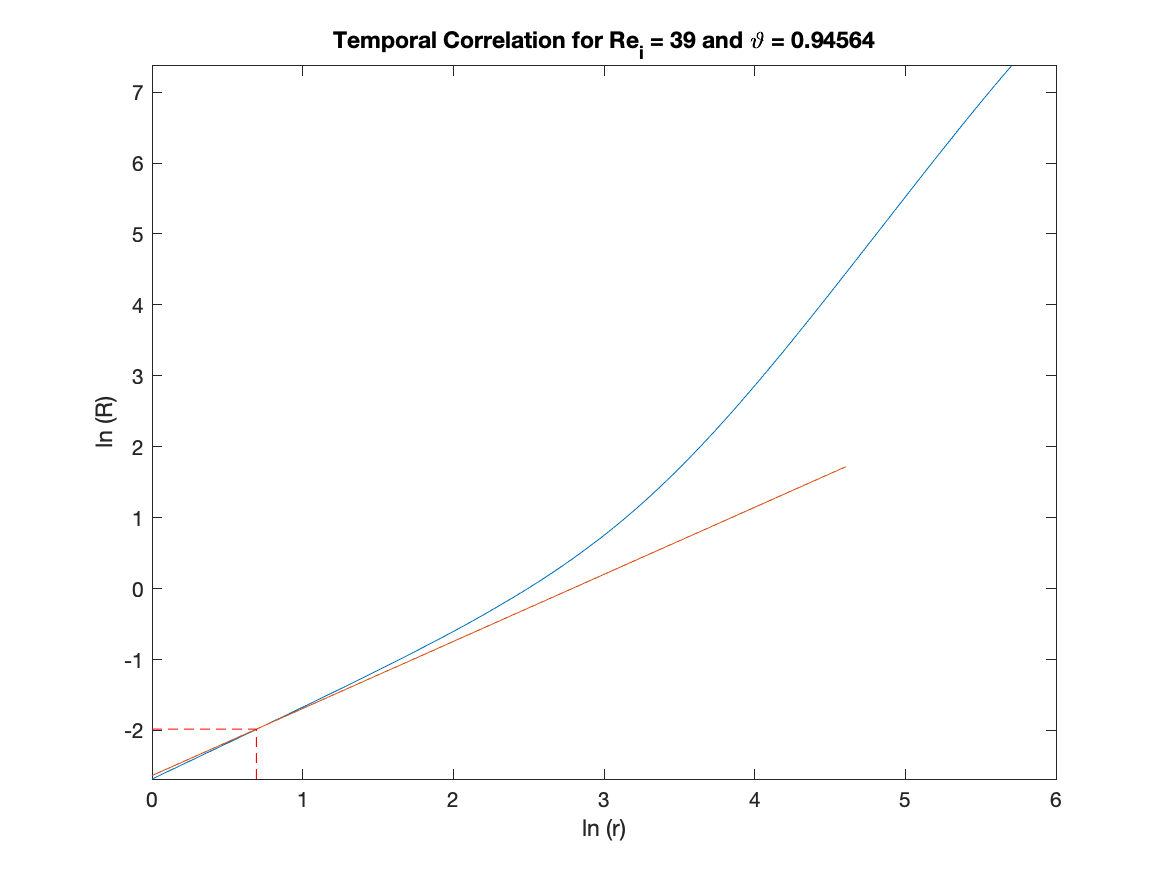}
	\end{subfigure}
	\begin{subfigure}[b]{0.45\linewidth}
		\includegraphics[width=\linewidth]{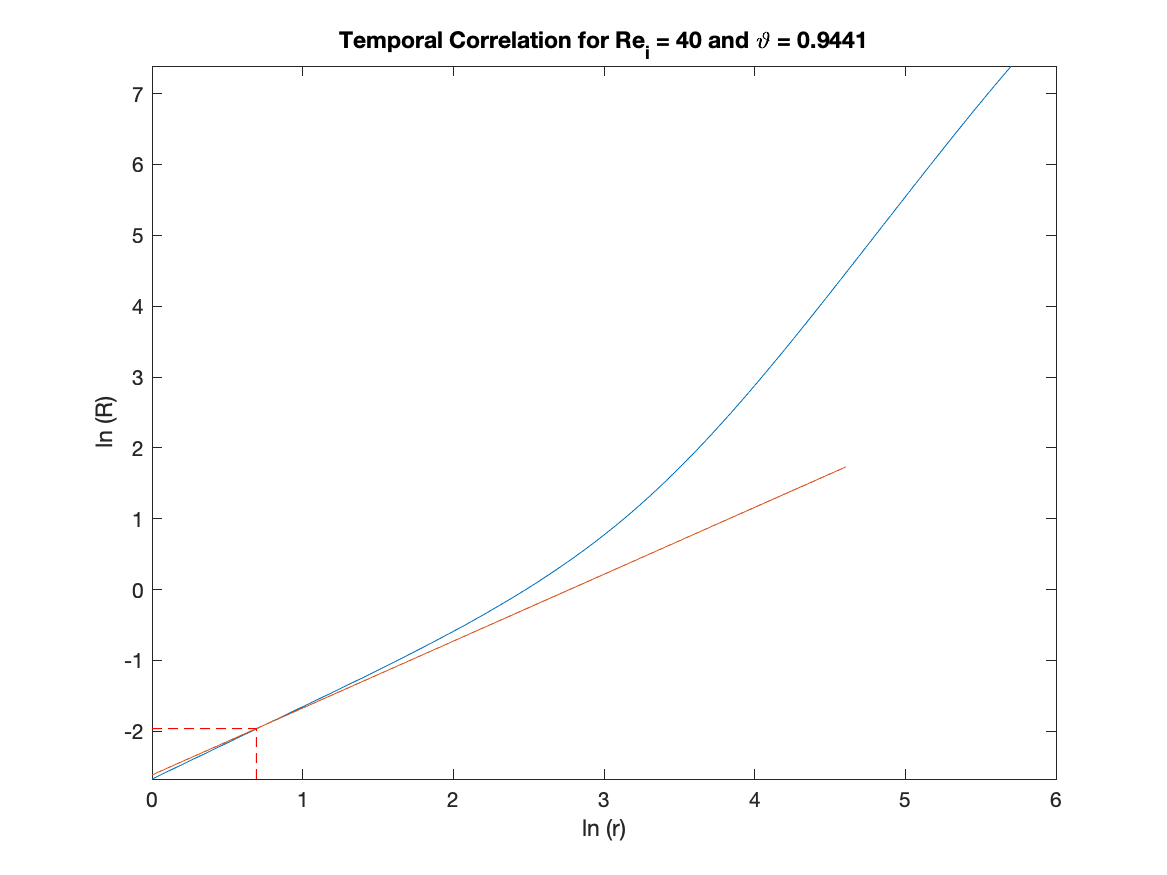}
	\end{subfigure}
	\begin{subfigure}[b]{0.45\linewidth}
		\includegraphics[width=\linewidth]{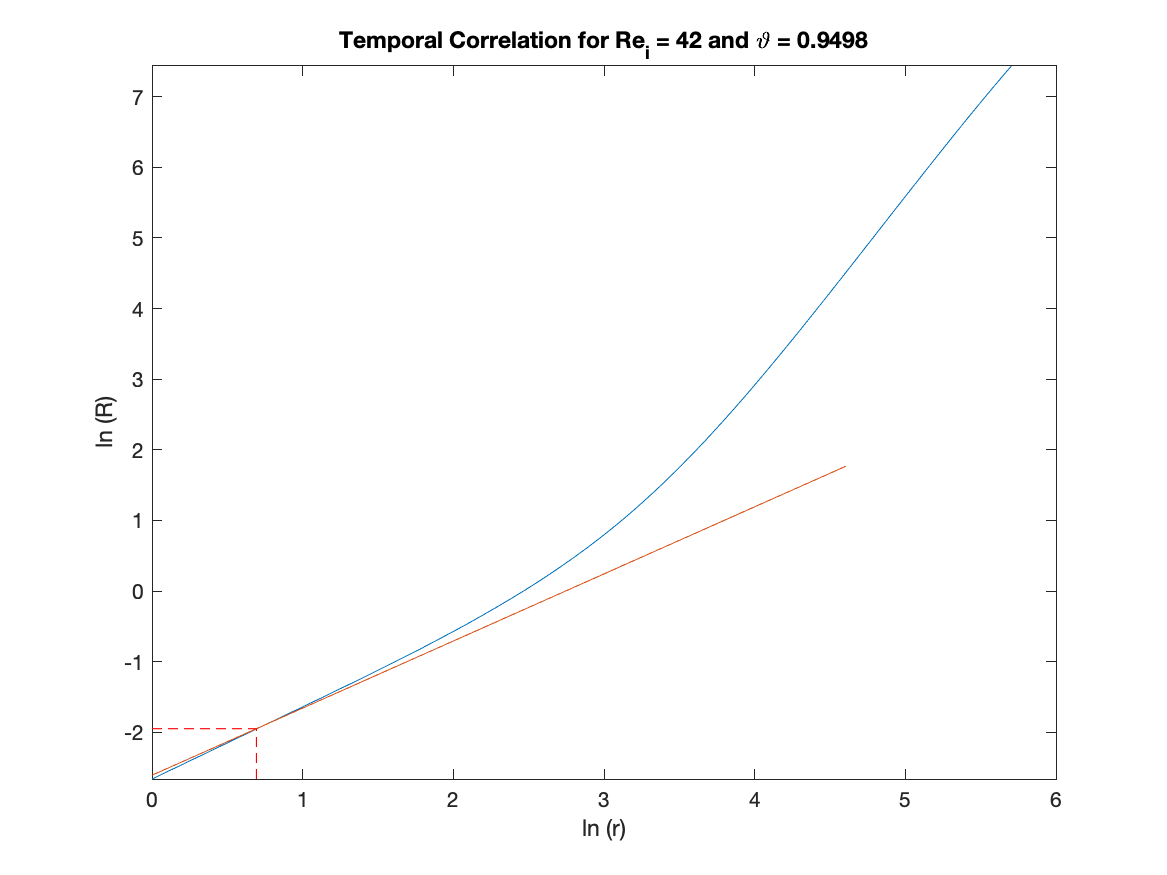}
	\end{subfigure}
	\begin{subfigure}[b]{0.45\linewidth}
		\includegraphics[width=\linewidth]{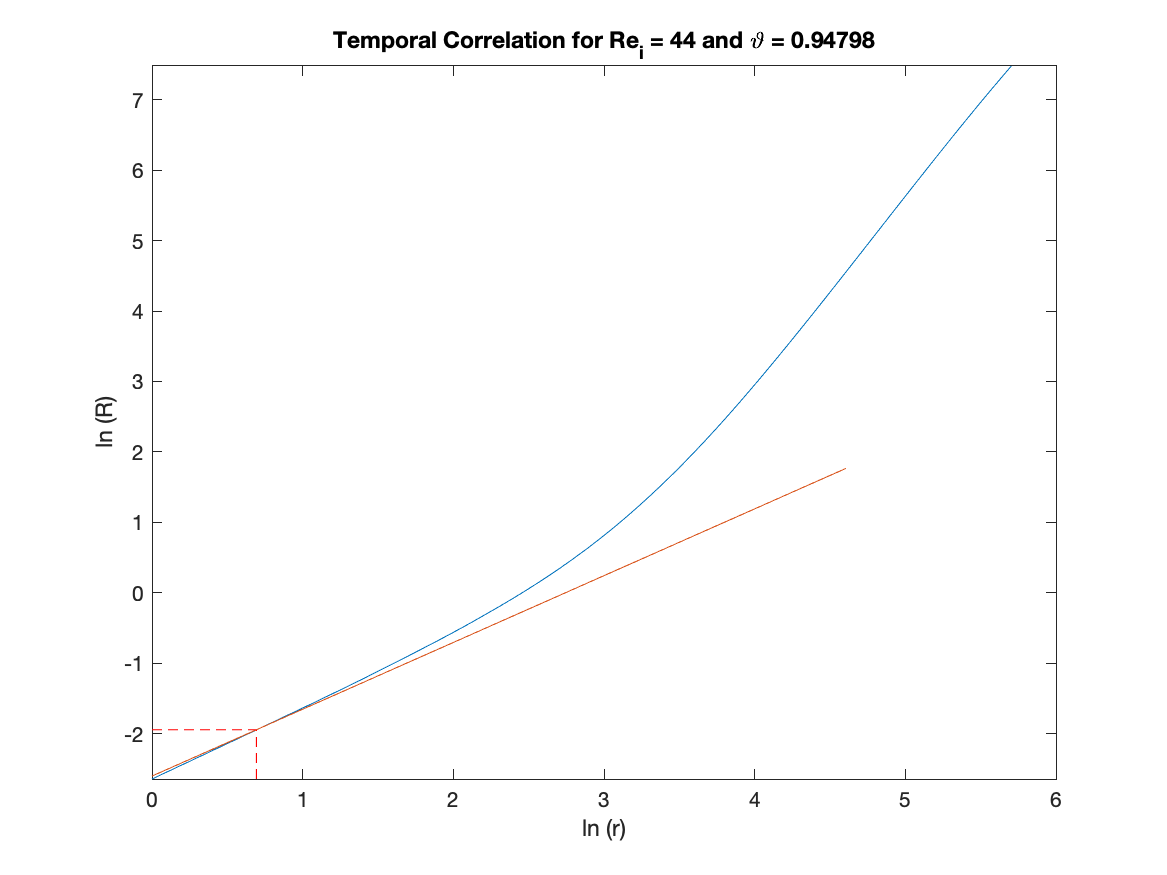}
	\end{subfigure}
	\caption{\enquote{Min} Case: plots showing the temporal correlation function for an extended range of $Re_i$ for a fixed $\epsilon$}
	\label{sc06}
\end{figure}

\begin{figure}[h!]
	\centering
	\begin{subfigure}[b]{0.45\linewidth}
		\includegraphics[width=\linewidth]{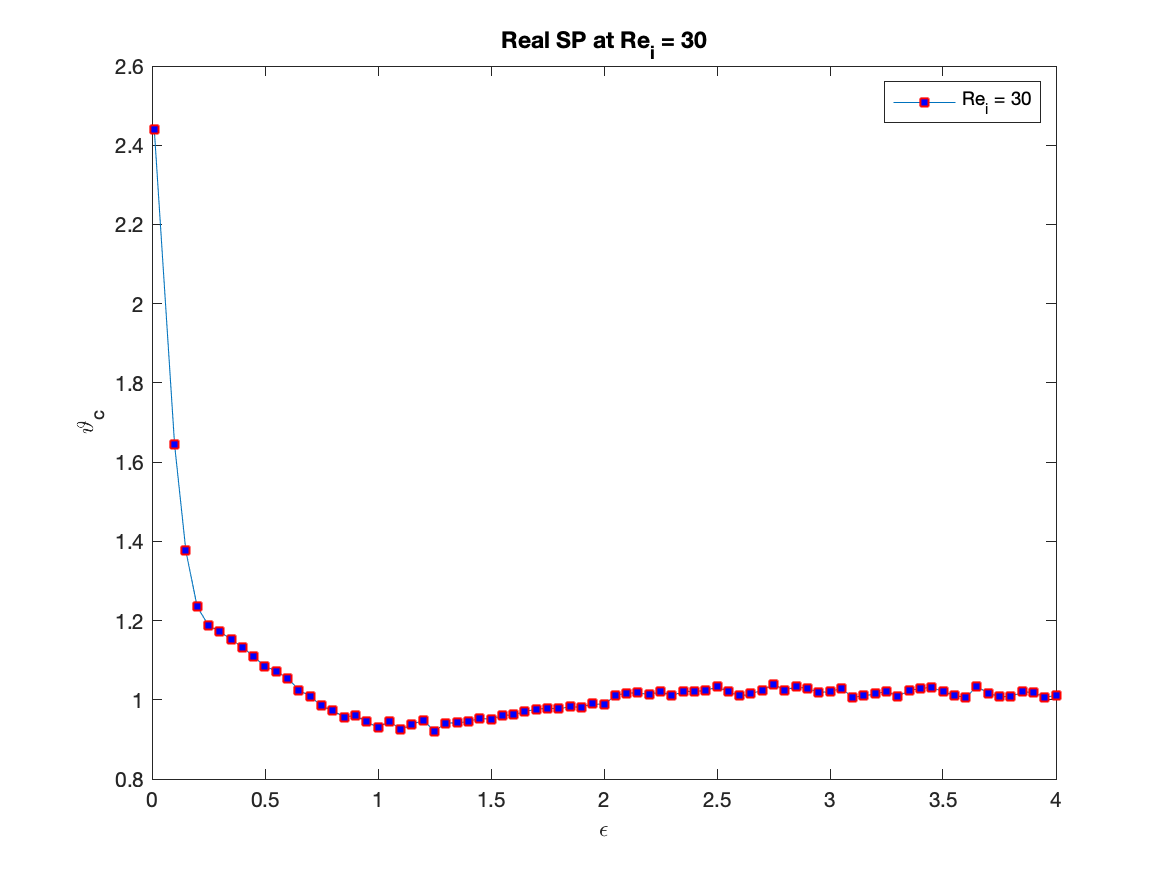}
	\end{subfigure}
	\begin{subfigure}[b]{0.45\linewidth}
		\includegraphics[width=\linewidth]{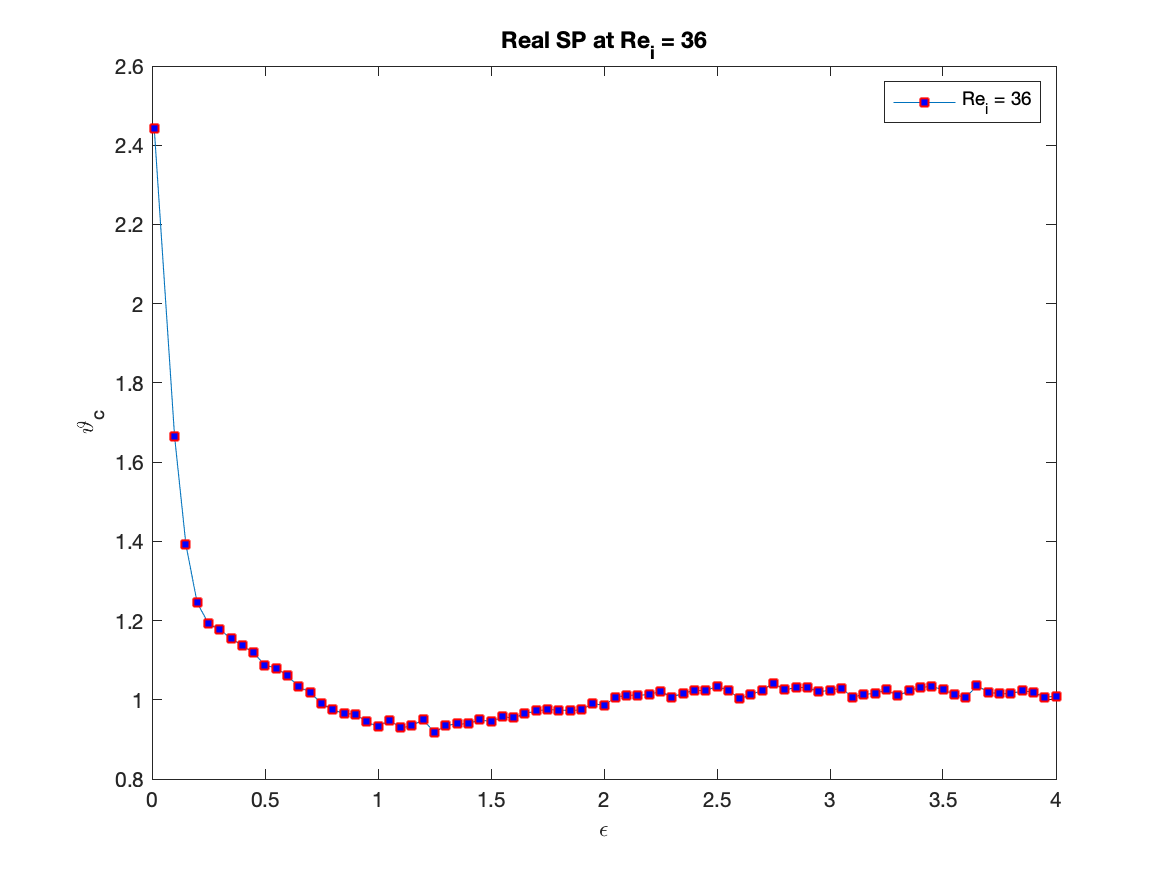}
	\end{subfigure}
	\begin{subfigure}[b]{0.45\linewidth}
		\includegraphics[width=\linewidth]{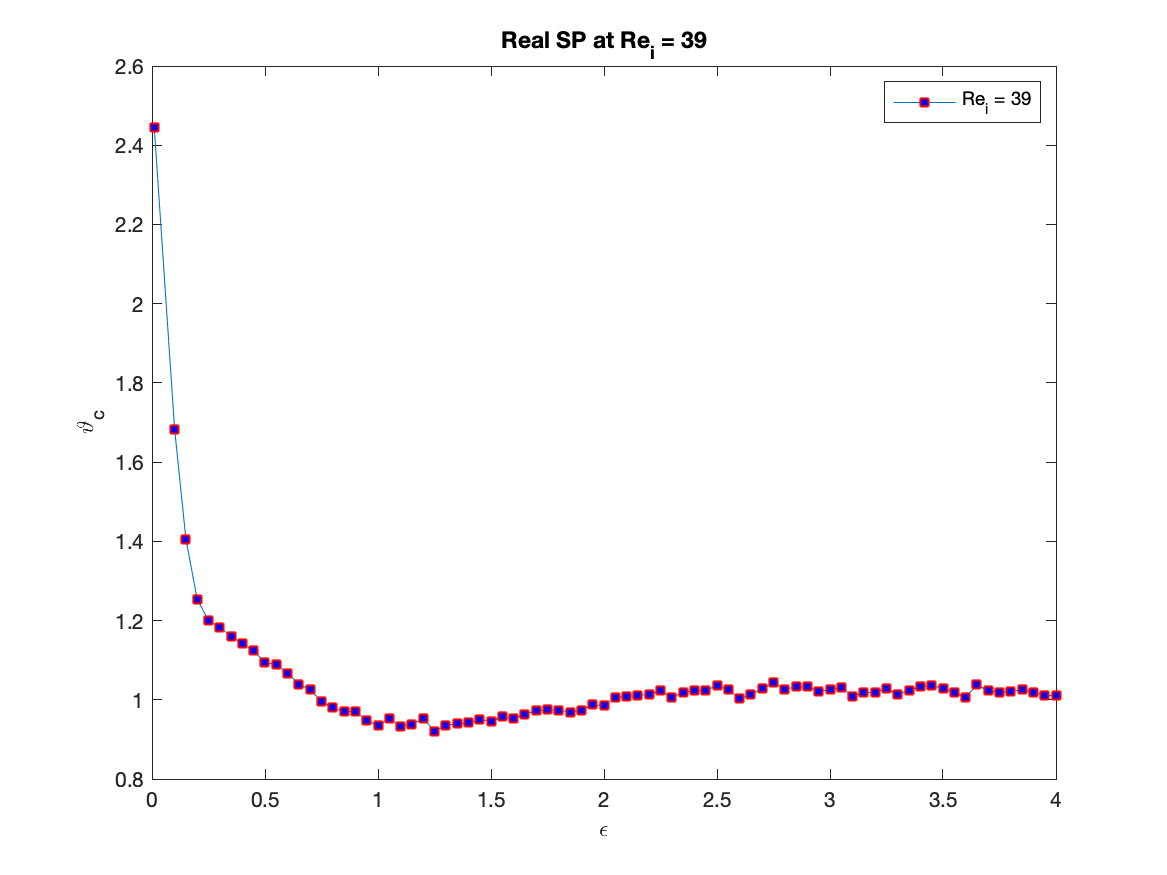}
	\end{subfigure}
	\begin{subfigure}[b]{0.45\linewidth}
		\includegraphics[width=\linewidth]{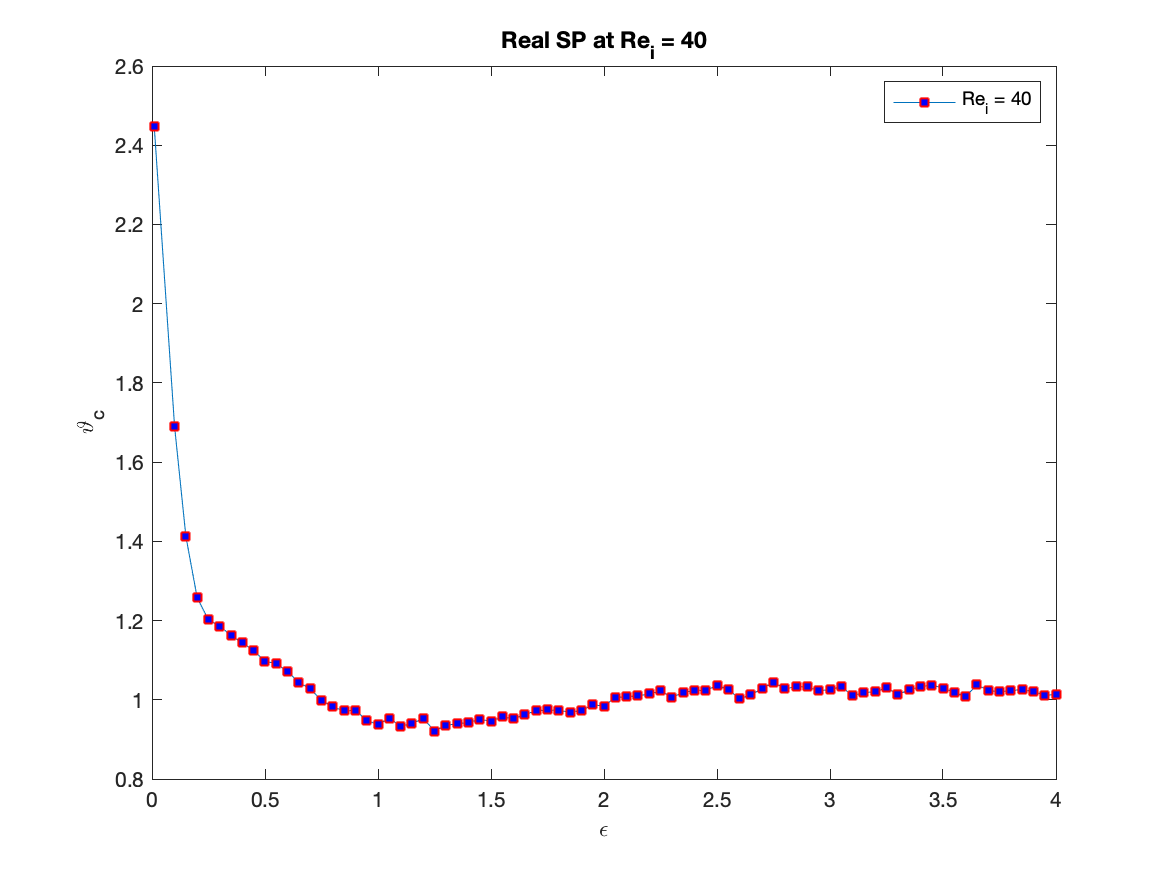}
	\end{subfigure}
	\begin{subfigure}[b]{0.45\linewidth}
		\includegraphics[width=\linewidth]{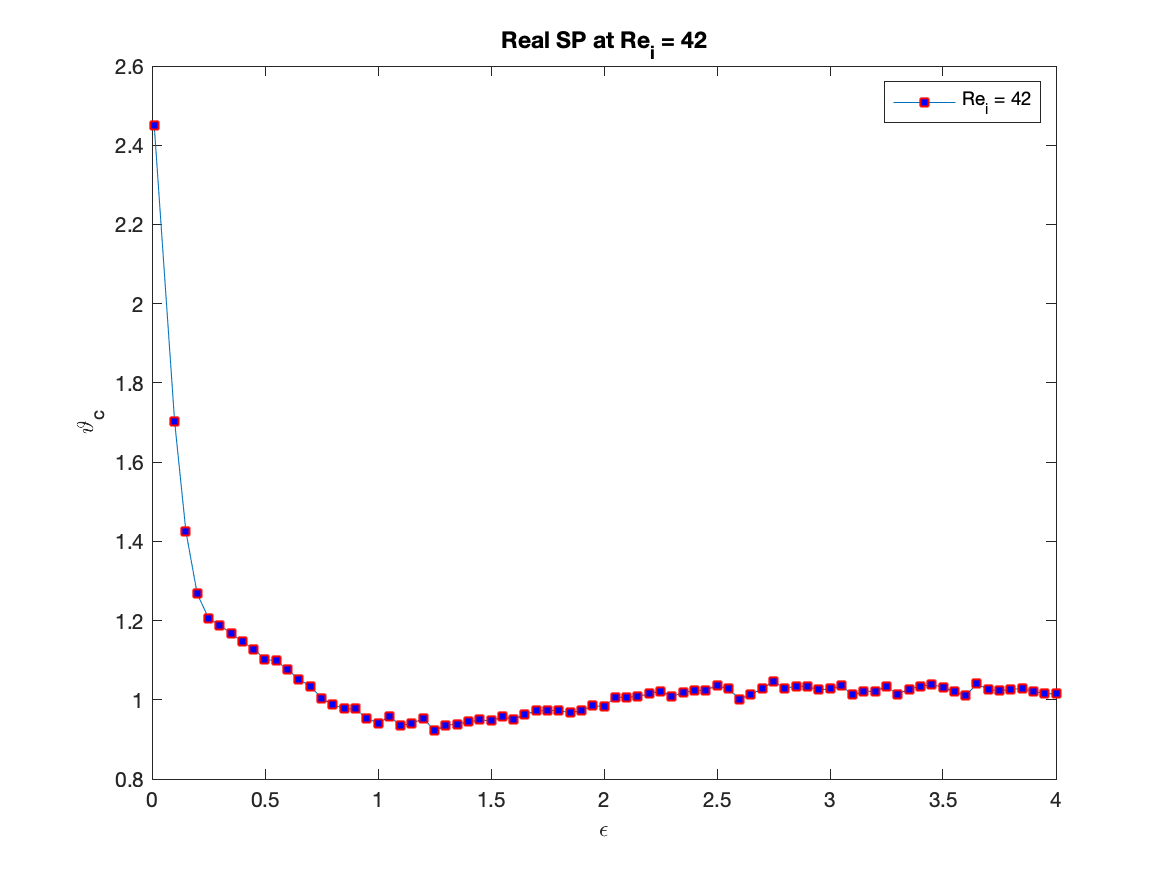}
	\end{subfigure}
	\begin{subfigure}[b]{0.45\linewidth}
		\includegraphics[width=\linewidth]{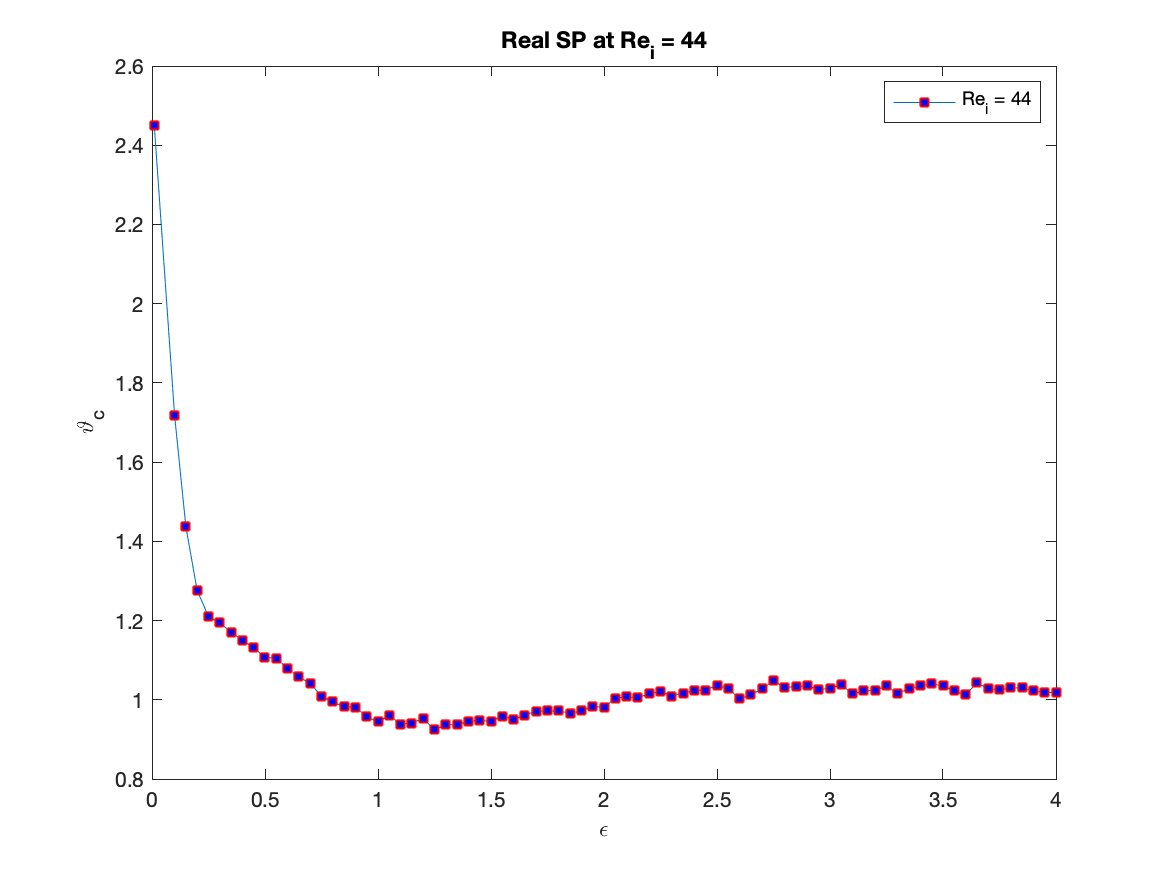}
	\end{subfigure}
	\caption{Plots showing the temporal correlation function for varius values of $Re_i$ for a fixed $\epsilon$, showing a maximum local at $\epsilon \sim 0.75$ and a local minimum at  $\epsilon \sim 0.6$}. 
	\label{sc05}
\end{figure}

The plots below show variations of the stability with increasing Reynolds number (Figure \ref{sc06}) and temporal correlation change with similarly varying Reynolds numbers (Figure \ref{sc05}).

\end{document}